\documentclass[twocolumn]{aastex62}

\usepackage{amsmath}

\shorttitle{Systematic Monitoring of LkCa 15}
\shortauthors{Sallum et al.}

\graphicspath{{./}{figures/}}

\usepackage{multirow}

\usepackage{enumitem}

\begin{document}

\title{Systematic Multi-Epoch Monitoring of LkCa 15: Dynamic Dust Structures on Solar-System Scales}

\author{Steph Sallum}
\affiliation{Department of Physics and Astronomy, University of California, Irvine, 4129 Frederick Reines Hall, Irvine, CA 92697, USA}

\author{Josh Eisner}
\affiliation{Steward Observatory, University of Arizona, 933 North Cherry Avenue, Tucson, AZ 85721, USA}

\author{Andy Skemer}
\affiliation{Department of Astronomy and Astrophysics, University of California, Santa Cruz, CA 95064, USA}

\author{Ruth Murray-Clay}
\affiliation{Department of Astronomy and Astrophysics, University of California, Santa Cruz, CA 95064, USA}

\begin{abstract}

We present the highest angular resolution infrared monitoring of LkCa 15, a young solar analog hosting a transition disk.
This system has been the subject of a number of direct imaging studies from the millimeter through the optical, which have revealed multiple protoplanetary disk rings as well as three orbiting protoplanet candidates detected in infrared continuum (one of which was simultaneously seen at H$\alpha$). 
We use high-angular-resolution infrared imaging from 2014-2020 to systematically monitor these infrared signals and determine their physical origin.
We find that three self-luminous protoplanets cannot explain the positional evolution of the infrared sources, since the longer time baseline images lack the coherent orbital motion that would be expected for companions.
However, the data still strongly prefer a time-variable morphology that cannot be reproduced by static scattered-light disk models. 
The multi-epoch observations suggest the presence of complex and dynamic substructures moving through the forward-scattering side of the disk at $\sim20$ AU, or quickly-varying shadowing by closer-in material. 
We explore whether the previous H$\alpha$ detection of one candidate would be inconsistent with this scenario, and in the process develop an analytical signal-to-noise penalty for H$\alpha$ excesses detected near forward-scattered light. 
Under these new noise considerations, the H$\alpha$ detection is not strongly inconsistent with forward scattering, making the dynamic LkCa 15 disk a natural explanation for both the infrared and H$\alpha$ data.

\end{abstract}

\keywords{transition disks - star and planet formation - LkCa 15 - high resolution imaging - interferometry - Facility: Keck:II (NIRC2), LBT (LMIRCam)}

\section{Introduction}\label{sec:intro}

Transition disks are protoplanetary disks with inner clearings first inferred from spectral energy distribution fitting \citep{1989AJ.....97.1451S}, and later confirmed in sub-millimeter imaging \citep[e.g][]{2011ApJ...732...42A}.
In addition to large clearings, some of these disks exhibit complex substructures \citep[e.g. gaps, warps, and spirals, in both large and small grains;][]{2018ApJ...869L..41A,2019ApJ...883..100S}, all of which can be connected to embedded protoplanets via hydrodynamical modeling \citep[e.g.][]{1999ApJ...514..344B,2006A&A...453.1129P,2018ApJ...862..103D}.
These results, along with transition disks' low stellar accretion rates \citep[e.g.][]{2015MNRAS.450.3559N}, have established these objects as natural planet formation laboratories \citep[e.g.][]{2019MNRAS.486..453L}. 
Indeed, transition disks have been targeted by a variety of direct imaging protoplanet searches \citep[e.g][]{2020A&A...633A.119Z}, with the first robust protoplanet detection recently made in the PDS 70 transition disk \citep[e.g.][]{2018A&A...617A..44K,2019NatAs...3..749H}.

\begin{figure}
\begin{center}
\includegraphics[width=0.9\columnwidth]{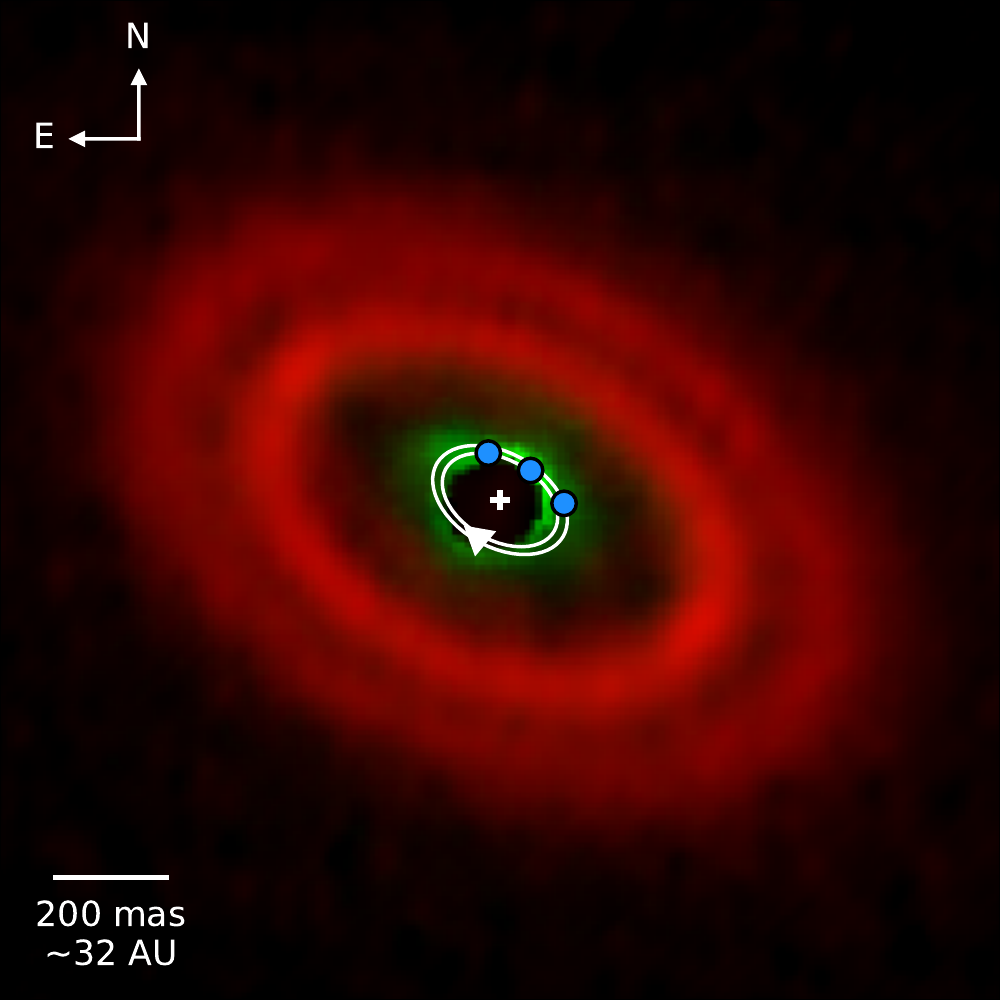}
\caption{Color composite image of selected, previously-published LkCa 15 datasets and companion candidate positions. Red shows ALMA 1.3 mm observations from \citet{2020A&A...639A.121F} that led to the identification of annular substructures. Green shows SPHERE J band polarized light imaging from \citet{2016ApJ...828L..17T}. Blue scattered points show the positions of the three companion candidates in \citet{2015Natur.527..342S}, which, when compared to the three companion signals in \citet{2012ApJ...745....5K}, exhibited apparent position angle changes along the orbits marked in white (in the direction indicated by the white arrow). The detection of polarized light due to scattering by disk material \citep{2016ApJ...828L..17T}, and follow-up observations that were inconsistent with three distinct companion orbits \citep{2019ApJ...877L...3C} suggested that a $\sim20$ AU, small-grain disk could be at least partially responsible for the companion candidate signals.\label{fig:composite}}
\end{center}
\end{figure}

One well studied transition disk system is LkCa 15 - a 1.2 $\mathrm{M_\odot}$, K5 T Tauri star in the Taurus star forming region at a distance of 159 pc \citep{2018A&A...616A...1G,2019MNRAS.483L...1D}. 
Early millimeter imaging of its circumstellar disk indicated an inner clearing with a radial extent of $\sim$40-50 AU \citep{2006A&A...460L..43P,2011ApJ...742L...5A,2011ApJ...732...42A}.
Recent ALMA data reveal more complex disk structure, with concentric rings located at approximately 42 AU, 69 AU, and 101 AU \citep[Figure \ref{fig:composite};][]{2020A&A...639A.121F}. 
The 42 AU ring exhibits a horseshoe-like morphology, with a clump and an arc potentially caused by dust trapping in the Lagrangian points of an undetected planet \citep{2022ApJ...937L...1L}. 
These features and azimuthal dust trapping in the adjacent 69 AU ring all point to ongoing planet formation in LkCa 15. 

LkCa 15 has been the subject of a number of direct imaging studies aimed at both searching for embedded accreting planets and characterizing small grain disk material. 
Keck non-redundant masking (NRM; Section \ref{sec:obs}) at K$^\prime$ ($2.12~\mu$m) and L$^\prime$ ($3.8~\mu$m) in 2009-2010 revealed the presence of multiple infrared point sources within the millimeter clearing \citep{2012ApJ...745....5K}.
The morphology and point source fluxes were explained as a single accreting protoplanet (the central component with bluer colors) surrounded by dusty material (two flanking components with redder colors). 
Later observations from the Large Binocular Telescope also revealed three sources  (b, c, d), all of which were detected at L$^\prime$, two (b, c) at K$\mathrm{s}$ ($2.16~\mu$m), and one (b) in H$\alpha$ (656.3 nm) spectral differential imaging from Magellan/MagAO \citep{2015Natur.527..342S}. 
When compared to astrometry from \citet{2012ApJ...745....5K}, the position angle evolution of the companions was consistent with Keplerian orbital motion with three distinct semimajor axes. 
The position angle evolution, combined with infrared and H$\alpha$ fluxes consistent with accretion, made three orbiting protoplanets a natural explanation for the observations. 

Several follow-up imaging studies called this protoplanet scenario into question, with both small-grain disk material and an H$\alpha$-bright disk wind as alternative explanations. 
VLT/SPHERE J band polarimetric imaging (Figure \ref{fig:composite}) revealed asymmetric scattering structures at the locations of the two L$^\prime$ companions not detected at H$\alpha$, suggesting that scattered light by disk material may at least partially account for the infrared NRM signals \citep{2016ApJ...828L..17T}. 
Spectro-astrometry from William Herschel Telescope later revealed no obvious signs of H$\alpha$-bright companions, but instead evidence for a symmetric morphology such as a disk wind \citep{2018A&A...618L...9M}. 
However, these H$\alpha$ observations could not rule out the presence of LkCa 15 b, since their sensitivity limits were within the 1$\sigma$ errors on b's contrast.

Imaging at K and L$^\prime$ from Subaru/SCExAO and Keck/NIRC2 was also used to investigate the orbital motion scenario \citep{2019ApJ...877L...3C}. 
This multi epoch imaging showed extended features that did not move according to the best-fit orbits published in \citet{2015Natur.527..342S}, but this study did not carry out independent fits to constrain peak position angles and/or search for variability. 
Most recently, a single epoch of 2.1-2.3 $\mu$m VLT/SPHERE NRM observations were shown to be better explained by a smooth extended structure than by the three companions \citep{2022ApJ...931....3B}.
This motivates the analysis of a broader, multi-band and multi-epoch NRM study to investigate the disk structure and planet detections. 

Here we present a systematic, multi-band and multi-epoch study of the inner regions of the LkCa 15 system.
Using LBT and Keck NRM observations from 2014-2020 at H, K$\mathrm{_s}$, and L$^\prime$ bands, we compare the disk and multi-companion scenarios. 
We apply a combination of model fitting and image reconstruction to characterize the system morphology at each observational epoch, searching for changes with time.
In Section \ref{sec:obs} we describe this broad dataset, and in Sections \ref{sec:datared} and \ref{sec:analysis} we discuss our data reduction and analysis strategies.
Section \ref{sec:res} describes the results of model fits and image reconstructions, which we discuss in the context of previous studies in Section \ref{sec:disc} before concluding in Section \ref{sec:conc}.

\begin{deluxetable*}{lcccccccccccl}
\tablecaption{Observations\label{tab:obs}}
\tablewidth{700pt}
\tabletypesize{\scriptsize}
\tablehead{
\colhead{Epoch$^*$} & \colhead{Date} & \colhead{Instrument} &
\colhead{Filter} & \colhead{t$_\mathrm{i}$} & 
\colhead{n$_\mathrm{f}$} & \colhead{n$_\mathrm{p}$} & 
\colhead{t$_\mathrm{total}$} & \colhead{$\Delta$PA} &
\colhead{Calibrators} & \colhead{Seeing} &  \colhead{Ref.$^\parallel$} \\ 
\colhead{} & \colhead{(yymmdd)} & \colhead{} & 
\colhead{}& \colhead{(s)} & \colhead{} & \colhead{} & 
\colhead{(h)} & \colhead{($^\circ$)} & \colhead{} & \colhead{$^{\prime\prime}$} & \colhead{} 
} 
\startdata
2014-12 L$^\prime$ & 141215 & LBTI/LMIRCam R$^\dagger$ & Std-L & 10 & 40 & 15 & 1.7 & 130 & HD 284581, HD 284668, GM Aur$^\ddagger$ & 0.76$\pm$0.09 & [1]\\
2014-12 L$^\prime$ & 141215 & LBTI/LMIRCam L$^\dagger$ & Std-L & 10 & 40 & 15 & 1.7 & 130 & HD 284581, HD 284668, GM Aur$^\ddagger$ & 0.76$\pm$0.09 & [1]\\
2015-02 $\mathrm{K_s}$ & 150205 & LBTI/LMIRCam R$^\dagger$ & Ks & 20 & 20 & 9 & 1.0 & 94 & HD 284581, HD 284668, GM Aur$^\ddagger$ & 0.92$\pm$0.25 & [1]\\
2015-02 $\mathrm{K_s}$ & 150205 & LBTI/LMIRCam L$^\dagger$ & Ks & 20 & 20 & 9 & 1.0 & 94 & HD 284581, HD 284668, GM Aur$^\ddagger$ & 0.92$\pm$0.25 & [1]\\
2015-02 $\mathrm{K_s}$ & 150207 & LBTI/LMIRCam R$^\dagger$ & Ks & 20 & 20 & 10 & 1.1 & 104 & HD 284581, HD 284668, GM Aur$^\ddagger$ & 0.93$\pm$0.21 & [1]\\
2015-02 $\mathrm{K_s}$ & 150207 & LBTI/LMIRCam R$^\dagger$ & Ks & 20 & 20 & 10 & 1.1 & 104 & HD 284581, HD 284668, GM Aur$^\ddagger$ & 0.93$\pm$0.21 & [1]\\
2016-02 L$^\prime$ & 160217 & LBTI/LMIRCam R$^\dagger$ & Std-L & 10 & 40 & 4 & 0.4 & 34 & HD 284581, HD 284668 & 1.00$\pm$0.11 & [2]\\
2016-02 L$^\prime$ & 160217 & LBTI/LMIRCam L$^\dagger$ & Std-L & 10 & 40 & 8 & 0.9 & 12 & HD 284581, HD 284668 & 1.00$\pm$0.11 & [2]\\
2016-02 L$^\prime$ & 160220 & LBTI/LMIRCam R$^\dagger$ & Std-L & 10 & 40 & 8 & 0.9 & 31 & HD 284581, HD 284668 & 0.87$\pm$0.07 & [2]\\
2016-02 L$^\prime$ & 160220 & LBTI/LMIRCam L$^\dagger$ & Std-L & 10 & 40 & 9 & 1.0 & 31 & HD 284581, HD 284668 & 0.87$\pm$0.07 & [2]\\
2016-11 L$^\prime$ & 161116 & LBTI/LMIRCam R$^\dagger$ & Std-L & 10 & 40 & 6 & 0.7 & 116 & HD 284581, HD 284668 & 0.71$\pm$0.08 & [2]\\
2016-11 L$^\prime$ & 161116 & LBTI/LMIRCam L$^\dagger$ & Std-L & 10 & 40 & 6 & 0.7 & 117 & HD 284581, HD 284668 & 0.71$\pm$0.08 & [2]\\
2018-01 L$^\prime$ & 171231 & Keck/NIRC2 & L$^\prime$ & 20 & 20 & 8 & 0.9 & 16 & HD 284581, HD 284668 & $\sim$0.5-0.6$^{\S}$ & --\\ 
2018-01 L$^\prime$ & 180101 & Keck/NIRC2 & L$^\prime$ & 20 & 20 & 11 & 1.2 & 180 & HD 284581, HD 284668 & $\sim$0.6-0.8$^{\S}$ & --\\
2018-11 L$^\prime$ & 181121 & Keck/NIRC2 & L$^\prime$ & 20 & 20 & 11 & 1.2 & 171 & HD 284581, HD 284668 & 0.65$\pm$0.11 & --\\
2019-01 H & 190110 & Keck/NIRC2 & H & 20 & 20 & 14 & 1.6 & 187 & HD 284581, HD 284668 & 0.76$\pm$0.27 &--\\
2019-01 H & 190111 & Keck/NIRC2 & H & 20 & 20 & 12 & 1.3 & 182 & HD 284581, HD 284668 & 0.51$\pm$0.12 & --\\
2020-01 L$^\prime$ & 200102 & Keck/NIRC2 & L$^\prime$ & 20 & 20 & 16 & 1.8 & 188 & HD 284581, HD 284668 & N/A$^{\S}$ & --
\enddata

\footnotesize{$^*$ During the analysis we combine multiple observing nights taken closely in time into individual ``epochs," which we then treat as single datasets during the model fitting and image reconstruction. We list the observational parameters for the individual nights here, but for the remainder of the paper we refer to the datasets using the epoch labels in the first column.}

\footnotesize{$^\parallel$ References for previously-published datasets that we re-reduce here: [1] \citet{2015Natur.527..342S}, [2] \citet{2016SPIE.9907E..0DS}. The -- symbol indicates new data obtained as part of this study.}

\footnotesize{$^\dagger$ We observed with the LBTI in single-aperture mode (without co-phasing), which places images from each 8-meter primary mirror in a different location on the detector. Since for some datasets poor adaptive optics performance led to data losses for just one of the primaries, we list observational parameters for the left (L) and right (R) LBT mirrors separately.}

\footnotesize{$^\ddagger$ As described in Section \ref{sec:datared}, GM Aur was used as a calibrator in \citet{2012ApJ...745....5K} and \citet{2015Natur.527..342S}. We explore reductions with and without GM Aur, since its status as a young star may make it a contaminated calibrator. Including it yielded results comparable to analyses that excluded it, in agreement with previous NRM non-detections in GM Aur \citep{2011ApJ...731....8K}}. The results shown here are for reductions that utilize GM Aur as a calibrator, since including it allows us to measure instrumental errors at higher cadence.

\footnotesize{$^\S$ The listed seeing data for Keck observations are CFHT DIMM measurements. For nights where DIMM measurements were not available, we instead list approximate CFHT model estimates. Neither of these estimates were available for the 200102 observing night.}\vspace{-10pt}
\end{deluxetable*}

 \section{Observations}\label{sec:obs}
We re-analyze previously-published observations of LkCa 15 from the Large Binocular Telescope (LBT), which were first presented in \citet{2015Natur.527..342S} and \citet{2016SPIE.9907E..0DS} (Table \ref{tab:obs}). 
These data were taken in 2014-2016 and consisted of both L$^\prime$ and $\mathrm{K_s}$ imaging.
We also add new Keck L$^\prime$ and H band observations obtained in 2017-2020.
Here we describe the observing strategy for both the previously-published and new datasets.

All of the observations utilized the technique of non-redundant masking (NRM), which turns a conventional telescope into an interferometric array via a pupil-plane mask \citep[e.g.][]{2000SPIE.4006..491T}.
It delivers moderate contrast on angular scales down to and even within the classical diffraction limit, offering a resolution boost of a factor of a few compared to traditional imaging \citep[e.g.][]{2019JATIS...5a8001S,2014ApJ...780..171G}.
NRM's resolution has enabled detailed studies of the close-in environments around distant young stars both with and without adaptive optics, resulting in the identification of complex disk structures \citep[e.g.][]{1999Natur.398..487T,2001Natur.409.1012T,2019ApJ...883..100S} and companions \citep[e.g.][]{2008ApJ...678L..59I,2021AJ....161...28S}, including the infrared companion candidates under study here \citep[e.g.][]{2012ApJ...745....5K}. 

We observed LkCa 15 with NRM at the LBT between 2014 and 2016 using LBTI/LMIRCam \citep[e.g.][]{2014SPIE.9148E..03B,2010SPIE.7735E..3HS} using the 12-hole aperture mask, which placed 6 holes over each 8-meter primary mirror. 
Between 2017 and 2020 we observed using Keck 2/NIRC2, with NIRC2 in its 9-hole masking configuration.
During each night we observed LkCa 15 through transit in order to accumulate parallactic angle evolution and fill in the $(u,v)$ plane.
Given the sparse Fourier coverage of the mask, this enables more robust aperture synthesis and thus image reconstruction and model fitting \citep[e.g.][]{2017JOSAA..34..904T}. 
We alternated between the science target and unresolved calibrator stars, which served as point-spread function references. 
We broke each observation up into $\mathrm{n_{p}}$ pointings to each object, each of which consisted of $\mathrm{n_{f}}$ frames with $\mathrm{t_{i}}$ integration times, for a total integration time of $\mathrm{t_{tot} = n_p \times n_f \times t_i}$. 
We dithered the images between the top and bottom halves of the detector to perform background subtraction. 
Table \ref{tab:obs} lists the parallactic angle coverage, number of pointings, and total integration time, as well as the mean and standard deviation of the seeing for each epoch and wavelength.

\section{Data Reduction}\label{sec:datared}
We re-reduced the previously-published LBT data and reduced the new Keck data using \texttt{SAMPy},\footnote{https://github.com/JWST-ERS1386-AMI/SAMpy} a well-tested pipeline that has been applied to NRM data from VLT, Magellan, LBT, Keck, and \textit{JWST} \citep[e.g.][]{2015ApJ...801...85S,2019ApJ...883..100S,2021AJ....161...28S,2022SPIE12183E..2MS}.
Here we describe the image-level calibrations, extraction of Fourier observables, and calibration of Fourier observables. 
We also describe changes from the original reduction of the 2014-2016 LBT data.
For a thorough description of the pipeline, we refer the reader to \citet{2017ApJS..233....9S} and \citet{2022SPIE12183E..2MS}. 

\subsection{Image Calibrations and Fourier Extraction}
We first perform flat fielding of all raw images.
We then carry out dark, bias, and sky subtraction by subtracting the median of one dither position from every image in the other dither position. 
For LBT data, we next account for LMIRCam's readout channel biases \citep[e.g.][]{2012SPIE.8446E..4FL}. 
We then crop each interferogram and perform bad pixel correction. 
Lastly, for LBT data we apply a distortion correction using \texttt{dewarp} \citep{2019ascl.soft07008S}, following the methodology described in \citep{2015A&A...576A.133M}.

We next Fourier transform the images, which show the interference fringes formed by the mask.
We use the filter bandpass, camera platescale, and mask hole locations to generate synthetic power spectra that we use to choose Fourier sampling coordinates. 
For each dataset, we compare these sampling coordinates to the time-averaged power spectra of the unresolved calibrator observations, to check for any mask misalignments (e.g. caused by inconsistent filter wheel rotation; mask flexure; or imperfect distortion correction).
Sampling the Fourier transformed images, we calculate closure phases - sums of Fourier phases around baselines that form triangles - and squared visibilities - powers associated with the different baselines.

\subsection{Fourier Calibrations and Error Estimation}
We take the \texttt{polycal} \citep[e.g.][]{2015Natur.527..342S,2019ApJ...883..100S} approach to calibrating the squared visibilities and closure phases.
We fit polynomial functions in time to the observables for each baseline and closing triangle. 
We then take the best-fit polynomials and sample them at the times of the science observations to estimate the systematic errors.
We subtract the systematic closure phases from the science closure phases, and divide the systematic squared visibilities into the science squared visibilities.
We explore a range of polynomial fits to the calibrators and adopt the one that, when sampled at the time of the science observations, minimizes the scatter in the calibrated science data.

For nearly all L$^\prime$ datasets $\mathrm{0^{th}-2^{nd}}$ order polynomials provide the best calibration.
For K$\mathrm{_s}$ and H band datasets $\mathrm{3^{rd}-5^{th}}$ orders minimize the scatter.
This suggests lower and/or more variable image quality for these observations.
This is to be expected given their shorter wavelengths and thus lower AO-corrected Strehls, particularly at H band.
This is also consistent with the higher and more variable seeing during the 2015 K$\mathrm{_s}$ observations (Table \ref{tab:obs}).

\subsection{Updates To 2014-2016 LBT Data Reduction}
A handful of changes were made to the pipeline between the LBT reductions published in \citet{2015Natur.527..342S} and \citet{2016SPIE.9907E..0DS} and those presented here. 
Distortion corrections were not applied to the original reductions; we now correct the raw images for LBT/LMIRCam's known distortion \citep[e.g.][]{2015A&A...579C...2M} using \texttt{dewarp} \citep{2019ascl.soft07008S}. 
We also use an updated prescription for the LBT mask hole locations, which results in a slightly better match between the Fourier sampling coordinates and the peak visibility amplitudes for each mask baseline. 
In \citet{2015Natur.527..342S}, an iterative calibration \citep[e.g.][]{2012ApJ...745....5K} was applied to the $\mathrm{K_s}$ data, calibrating toward the companion positions observed at L$^\prime$ for the same observational epoch. 
We do not apply this iterative calibration, opting instead for just the \texttt{polycal} method \citep[][]{2015Natur.527..342S}.

Beyond the reduction and calibration differences described above, we include the following higher-level analytical updates.
In \citet{2015Natur.527..342S}, the young star GM Aur was used as a calibrator for LkCa 15. 
We explore calibrations both with and without GM Aur, since it being a young star makes the possibility of contamination more likely. 
We find no significant difference between the quality of the calibration with and without GM Aur, which is consistent with previous NRM non-detections for that object \citep[e.g.][]{2011ApJ...731....8K}. 
We present the reduction that includes it as a calibrator, since it enables us to sample the instrumental errors at higher cadence. 
We also use a different image reconstruction algorithm from previous studies, as described in Section \ref{sec:imrecon}.

\section{Analysis}\label{sec:analysis}
When analyzing the data we combine observations taken closely in time (e.g. during adjacent nights) into single datasets for a given epoch.
As shown in Table \ref{tab:obs}, we designate these observational epochs according to the year, month, and bandpass.
This results in one H band epoch (2019-01 H), one $\mathrm{K_s}$ epoch (2015-02 $\mathrm{K_s}$), and six L$^\prime$ epochs (2014-12 L$^\prime$, 2016-02 L$^\prime$, 2016-11 L$^\prime$, 2018-01 L$^\prime$, 2018-11 L$^\prime$, and 2020-01 L$^\prime$).
For the remainder of the paper we refer to the data using those designations. 
For some of the analysis we also explore the time-averaged and/or multi-epoch properties of LkCa 15 by combining all of these L$^\prime$ datasets. In those cases we refer to this combined dataset as All L$^\prime$.

\subsection{Image Reconstruction}\label{sec:imrecon} 
Previous reconstructions of the 2015 and 2016 LBT datasets, published in \citet{2015Natur.527..342S} and \citet{2016SPIE.9907E..0DS}, respectively, were generated with BSMEM \citep{1994IAUS..158...91B}. 
However, depending on the choice of regularization, BSMEM is prone to over-resolve extended structures into multiple point-like sources \citep{2017ApJS..233....9S}. 
We therefore present images reconstructed with SQUEEZE \citep{2010SPIE.7734E..2IB}, an algorithm that uses Markov-Chain Monte Carlo methods to sample the image posterior.
SQUEEZE has been demonstrated to produce smoother reconstructions than BSMEM \citep[e.g.][]{2017ApJS..233....9S}. 
We note that when the BSMEM regularization parameter is chosen using the ``L-curve" method \citep[e.g.][]{Hansen1992}, the two algorithms produced comparable reconstructed images.

\subsection{Geometric Modeling} 
While image reconstruction has the advantage of being model-independent, the incomplete Fourier coverage of the mask makes it an under-constrained problem. 
Reconstructed images thus cannot be completely faithful to the underlying source brightness distribution. 
Furthermore, minimization and regularization methods specific to each algorithm have unique systematic effects on reconstructed image morphologies \citep[e.g.][]{2017JOSAA..34..904T}. 
We thus fit geometric models to the Fourier observables in addition to reconstructing images.

We explore two classes of models: (1) multiple point source models and (2) polar Gaussian ring models.
These geometric models are simply analytic representations of the true source morphology, with the multiple point source models representing ``clumpier" morphologies than the smooth polar Gaussian ring models.
For both model classes, we fit the datasets in two different ways to assess whether the multi-epoch observations require a static or dynamic scenario: (1) we fit each observational epoch and bandpass individually, and (2) we perform a geometric fit to all epochs simultaneously for each wavelength.

While these simplified models cannot capture the full complexity of the source brightness distribution, they enable the estimation of uncertainties and confidence intervals that are required for hypothesis testing.
Furthermore, comparisons of the dynamic and static model fits enable controlled searches for variability via goodness-of-fit metrics. 
Simple geometric models also allow us to understand the systematic effects of the image reconstructors via simulations (Section \ref{sec:imrecon_mod}). 
Below we describe the two model prescriptions.

\begin{figure*}
\includegraphics[width=\textwidth]{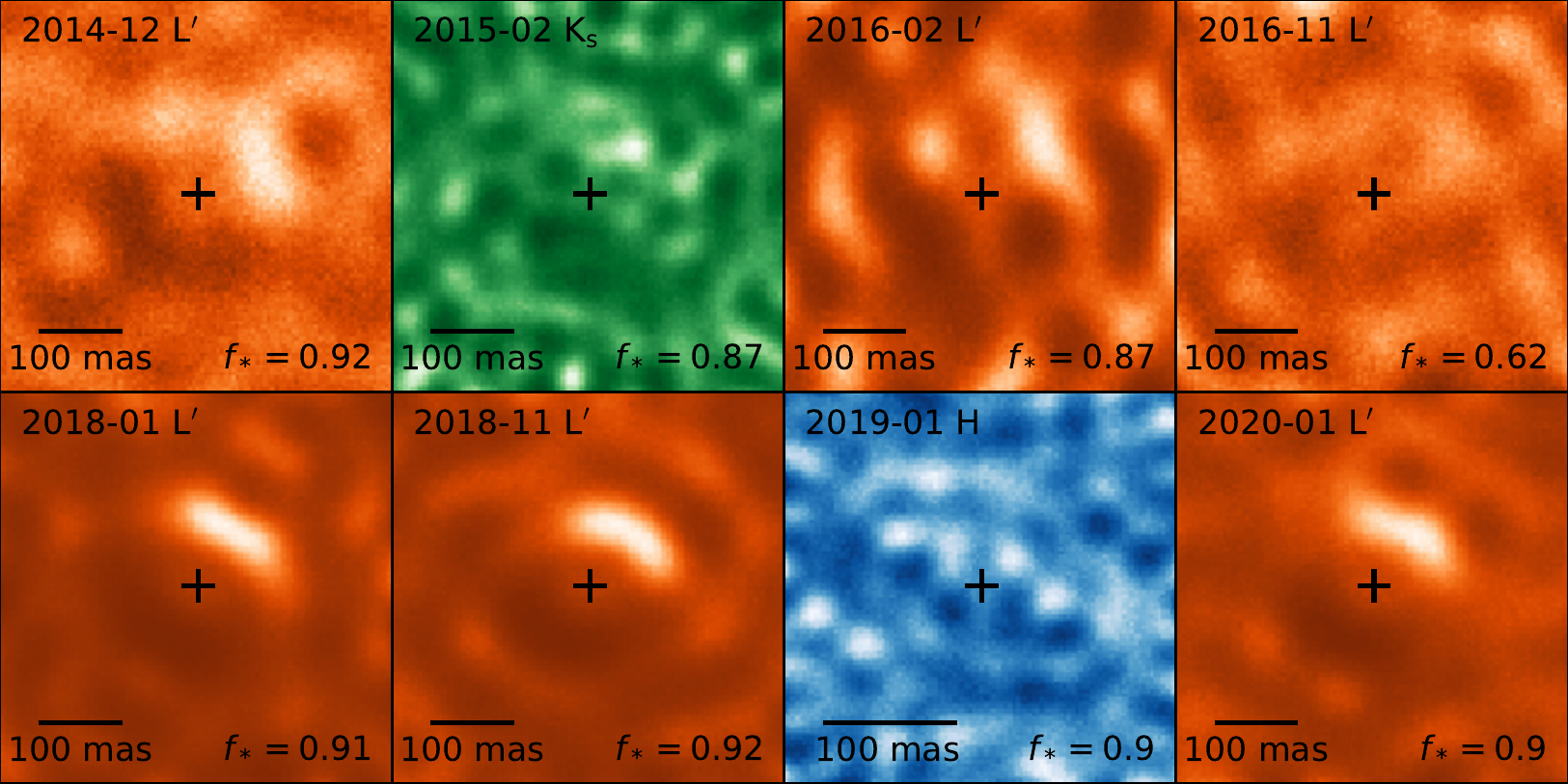}
\caption{SQUEEZE reconstructed images from individual epochs listed in Table \ref{tab:obs}. Each reconstruction includes an unresolved central model component with fractional flux $f_*$ that represents the star, with a location marked by the $+$. The panel color indicates the observing wavelength.\label{fig:recons_epochs}}
\end{figure*} 

\subsubsection{Multiple Point Source Models}
The multiple point source models consist of a central unresolved source (representing the star) plus up to three additional unresolved sources.
We refer to these models as one-point-source, two-point-source, and three-point-source models (referring to the number of sources in addition to the central one representing the star). 
For each of these point sources we define a separation $s_i$ measured in arcseconds, a position angle $\mathrm{PA_i}$ measured in degrees east of north, and a contrast $\Delta_i$ with respect to the central source measured in magnitudes. 

We model all point sources as delta functions, and calculate the analytic Fourier transform of the set of delta functions to generate the model squared visibilities and closure phases.
We use the python package \texttt{emcee} \citep{2013PASP..125..306F} to sample the parameter space with Markov-Chain Monte Carlo methods.
To avoid selecting for local likelihood maxima, we run \texttt{emcee} in parallel-tempering mode, which allows for multiple chains at different temperatures that can exchange information. 
We use 100 walkers and 10 temperatures for each fit, to ensure that the parameter space is explored efficiently.

\subsubsection{Polar Gaussian Ring Models}
To explore smoother source morphologies, we fit polar Gaussian ring models to the data using the same prescription as \citet{2022ApJ...931....3B}.
These allow for an inner and outer ring in addition to a central unresolved source (representing the star).
Each polar Gaussian ring has the following brightness distribution which allows for a finite radial extent as well as azimuthal asymmetry:
\begin{equation}
    I(r,\theta) = I_0 \exp{\left(-\frac{(r-r_0)^2}{2\sigma_r^2} - \frac{(\theta-\theta_0)^2}{2\sigma_\theta^2}\right)},
\end{equation}
where $I_0$ represents the peak flux in the ring, $r_0$ is the radius of the peak flux, $\theta_0$ is the position angle (measured E of N) of the peak flux, and $\sigma_r$ and $\sigma_\theta$ determine the radial and azimuthal extents of the ring, respectively. 

We also allow each ring to trace out an elliptical path, defined by an axis ratio ($r_a$) and a major axis position angle (measured E of N; $\phi_m$). 
Following \citet{2022ApJ...931....3B}, to make the ring geometries consistent with expectations for forward scattering by disk rims, we enforce the following relationship between the major axis position angle $\phi_m$ and the peak flux position angle $\theta_0$: 
\begin{equation}
    \theta_0 = \phi_m - 90^\circ.
\end{equation}
The axis ratio can be related to inclination ($i$) geometrically by the following relation:
\begin{equation}
i = \arccos\left({r_a}\right).
\end{equation}

When generating the model images, we create each ring brightness distribution using the above prescription, and then normalize the inner and outer rings relative to one another and to the central star. 
We parameterize the fractional flux of each component using $f_*,~f_o,~\mathrm{and}~f_i,$ to represent the star, outer ring, and inner ring, respectively. 
This amounts to two free parameters in the fit, since we apply the following prior that forces the total flux to be normalized to 1:
\begin{equation}
f_* + f_o + f_i = 1.
\end{equation}
Following \citet{2022ApJ...931....3B} we enforce a prior that requires the two inclinations and major axis position angles to be within 5$^\circ$ and 10$^\circ$ of each other, respectively. 
We also apply a prior to keep the radius of the outer ring larger than the inner ring radius, and we restrict the outer radius to be $ > 200$ mas.
As for the multiple point source models, for each fit we run \texttt{emcee} in parallel tempering mode with 100 walkers and 10 temperatures. 
Lastly, for completeness we also explore models consisting of only a single ring component, to assess the NRM observations' sensitivity to the $\sim50$ AU scattered light disk component detected in previous observations \citep[e.g.][]{2014A&A...566A..51T,2022ApJ...931....3B}.

\begin{figure*}
\begin{center}
\includegraphics[width=0.75\textwidth]{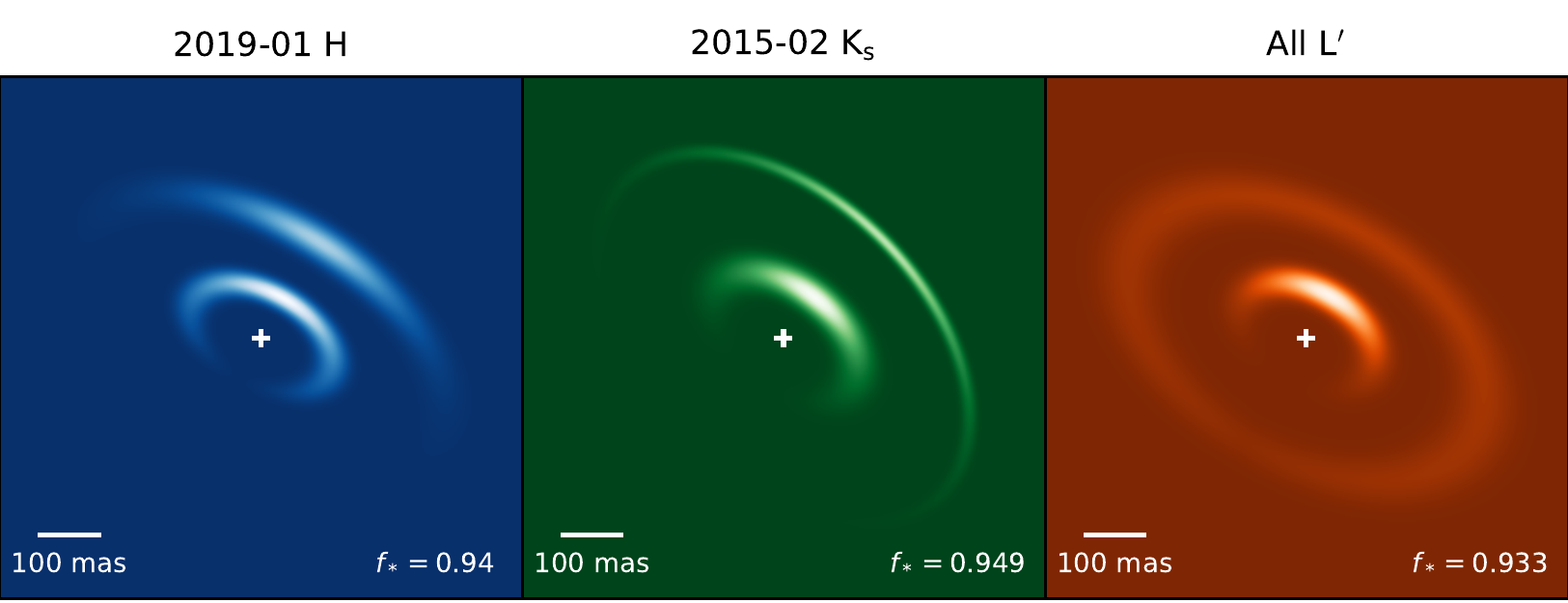}
\end{center}
\caption{Best-fit polar Gaussian ring models fit to all data obtained at H band (left), K$\mathrm{_s}$ band (center), and L$^\prime$ band (right). In each panel the location of the central star is marked with a $+$ and the star accounts for fractional flux $f_*$. The top three rows of Table \ref{tab:diskfits} list the best fit parameters for each dataset.\label{fig:2disk_bands}}
\end{figure*}

\begin{figure}
\vspace{15pt}
\includegraphics[width=\columnwidth]{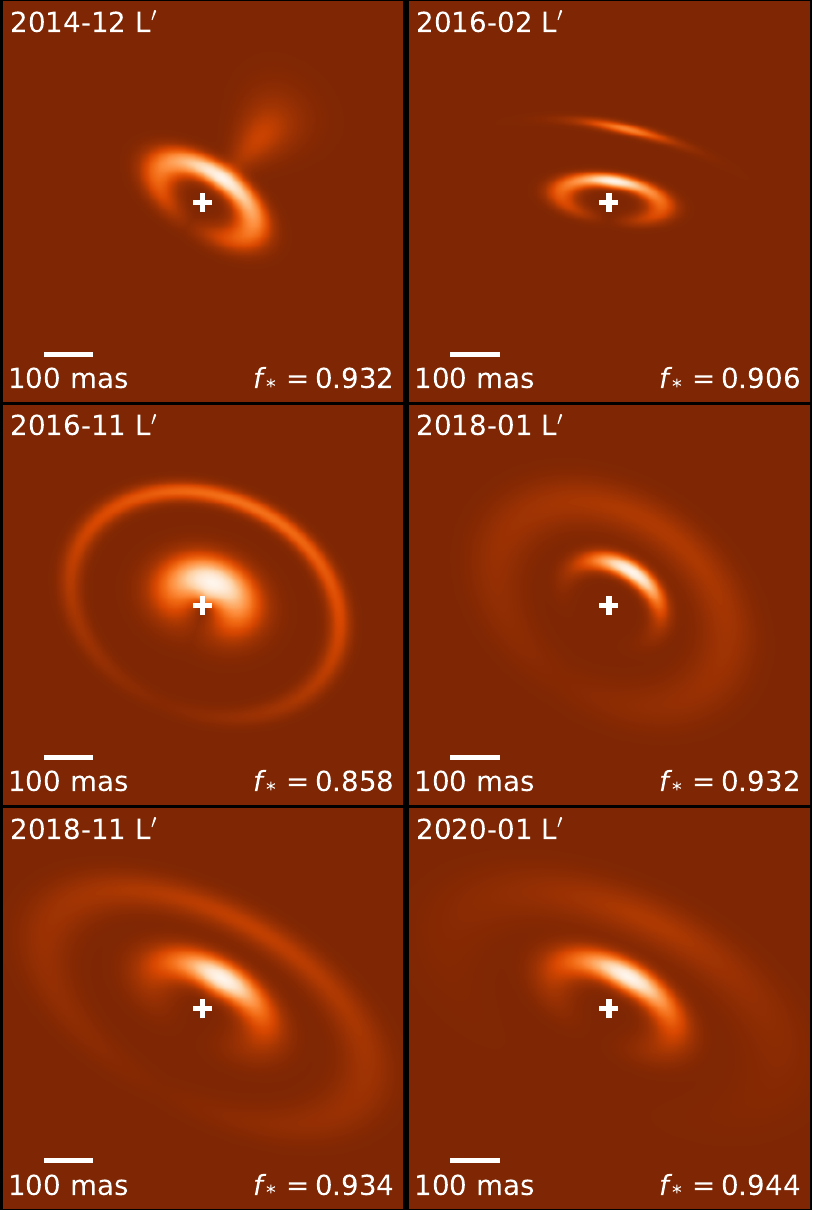}
\caption{Best-fit polar Gaussian ring models for each individual L$^\prime$ epoch. In each panel the location of the central star is marked with a $+$ and the star accounts for fractional flux $f_*$. Table \ref{tab:diskfits} lists the best fit parameters for each dataset.\label{fig:2disk_epochs}}
\end{figure}

\subsection{Image Reconstructions of Geometric Models}\label{sec:imrecon_mod}
In addition to standard model selection approaches (e.g. $\chi^2$ and Bayesian evidence comparisons) we reconstruct images from geometric models to assess their ability to explain the multi-epoch observations. 
For these tests we take a best-fit model of interest and generate its model squared visibilities and closure phases using the same (u,v) coverage and sky rotation as the data. 
We then reconstruct images in two different ways: (1) without adding noise to the model observables, and (2) adding enough Gaussian noise so that the scatter in the noisy model observables matches the observed scatter in the LkCa 15 Fourier observables. 
We then run SQUEEZE with identical reconstruction settings to those used on the real observations.

\section{Results}\label{sec:res}
\subsection{Multi-Epoch Reconstructed Images}
Figure \ref{fig:recons_epochs} shows the reconstructed images from each epoch. 
The 2014-12 L$'$ reconstruction shows a similar structure to that published in \citet{2015Natur.527..342S}, but without the appearance of three distinct point-like components. 
This is a result of the choice of image reconstruction algorithm; SQUEEZE is known to produce smoother reconstructions than BSMEM \citep[e.g.][]{2017ApJS..233....9S}. 
The 2016-02 L$^\prime$ and 2016-11 L$^\prime$ reconstructions are poor, owing to the high scatter in the closure phases and squared visibilities and the small amount of parallactic angle evolution (Table \ref{tab:obs}). 
In general the extended emission is seen at a consistent location - in an arc to the northwest of the star - across the L$'$ images taken between 2014 and 2020.
Differences exist in the extent of the arc and the peak brightness position angle from epoch to epoch.
The Ks and H band imaging also shows an arc-like feature at roughly the same position angle as the L$'$ emission, but with slightly smaller angular size.

\begin{deluxetable*}{lccccccccc}
\tablecaption{Multiple Point Source Fit Results\label{tab:compfits}}
\tablewidth{700pt}
\tabletypesize{\scriptsize}
\tablehead{
\colhead{Fit} & \colhead{$\mathrm{PA}_1$} & \colhead{$s_1$} & \colhead{$\Delta_1$} & \colhead{$\mathrm{PA}_2$} & \colhead{$s_2$} & \colhead{$\Delta_2$} & \colhead{$\mathrm{PA}_3$} & \colhead{$s_3$} & \colhead{$\Delta_3$} \\
\colhead{(1)} & \colhead{(2)} & \colhead{(3)} & \colhead{(4)} & \colhead{(5)} & \colhead{(6)} & \colhead{(7)} & \colhead{(8)} & \colhead{(9)} & \colhead{(10)}
} 
\startdata
    \multirow{2}{*}{2014-12 L$^\prime$ - 1PS} & $-74.2$ &  $0.073$ &  $5.1$ &   - &   - &   - &   - &   - &   -  \\
    & $\pm^{4.3}_{4.1}$ &  $\pm^{0.004}_{0.002}$ &  $\pm^{0.13}_{0.12}$ &   - &   - &   - &   - &   - &   -  \\ \hline
    \multirow{2}{*}{2014-12 L$^\prime$ - 2PS} & $-83.3$ &  $0.073$ &  $4.88$ &  $-20.1$ &  $0.077$ &  $5.22$ &   - &   - &   - \\
    &  $\pm^{3.6}_{2.6}$ &  $\pm^{0.004}_{0.002}$ &  $\pm^{0.11}_{0.09}$ &  $\pm^{5.1}_{5.4}$ &  $\pm^{0.01}_{0.005}$ &  $\pm^{0.14}_{0.15}$ &   - &   - &   - \\ \hline
    \multirow{2}{*}{2014-12 L$^\prime$ - 3PS} & $-86.4$ &  $0.079$ &  $5.15$ &  $-29.7$ &  $0.077$ &  $5.1$ &  $28.1$ &  $0.108$ &  $5.79$ \\
    & $\pm^{4.1}_{3.7}$ &  $\pm^{0.012}_{0.006}$ &  $\pm^{0.18}_{0.21}$ &  $\pm^{4.6}_{4.6}$ &  $\pm^{0.008}_{0.005}$ &  $\pm^{0.14}_{0.14}$ &  $\pm^{7.3}_{6.5}$ &  $\pm^{0.01}_{0.019}$ &  $\pm^{0.2}_{0.23}$ \\ \hline\hline
    \multirow{2}{*}{2015-02 $\mathrm{K_s}$ - 1PS} & $-43.6$ &  $0.078$ &  $5.41$ &   - &   - &   - &   - &   - &   - \\
    & $\pm^{1.6}_{1.2}$ &  $\pm^{0.002}_{0.002}$ &  $\pm^{0.09}_{0.06}$ &   - &   - &   - &   - &   - &   - \\ \hline
    \multirow{2}{*}{2015-02 $\mathrm{K_s}$ - 2PS} & $-79.9$ &  $0.112$ &  $5.87$ &  $-47.8$ &  $0.079$ &  $5.22$ &   - &   - &   - \\
    & $\pm^{1.1}_{1.1}$ &  $\pm^{0.003}_{0.003}$ &  $\pm^{0.1}_{0.11}$ &  $\pm^{1.3}_{1.4}$ &  $\pm^{0.001}_{0.001}$ &  $\pm^{0.09}_{0.06}$ &   - &   - &   - \\ \hline
    \multirow{2}{*}{2015-02 $\mathrm{K_s}$ - 3PS} & $-81.3$ &  $0.111$ &  $5.73$ &  $-47.7$ &  $0.079$ &  $5.2$ &  $-13.8$ &  $0.1$ &  $5.81$ \\
    & $\pm^{0.8}_{0.9}$ &  $\pm^{0.002}_{0.002}$ &  $\pm^{0.1}_{0.09}$ &  $\pm^{1.2}_{1.3}$ &  $\pm^{0.001}_{0.001}$ &  $\pm^{0.08}_{0.06}$ &  $\pm^{1.4}_{1.6}$ &  $\pm^{0.002}_{0.002}$ &  $\pm^{0.11}_{0.09}$ \\ \hline\hline
    \multirow{2}{*}{2016-02 L$^\prime$ - 1PS} & $-51.7$ &  $0.089$ &  $4.93$ &   - &   - &   - &   - &   - &   - \\
    & $\pm^{1.5}_{1.5}$ &  $\pm^{0.003}_{0.004}$ &  $\pm^{0.04}_{0.04}$ &   - &   - &   - &   - &   - &   -\\ \hline
    \multirow{2}{*}{2016-02 L$^\prime$ - 2PS} & $-59.9$ &  $0.084$ &  $4.84$ &  $-11.4$ &  $0.132$ &  $4.91$ &   - &   - &   -  \\
    & $\pm^{1.7}_{1.4}$ &  $\pm^{0.004}_{0.005}$ &  $\pm^{0.07}_{0.08}$ &  $\pm^{1.6}_{1.2}$ &  $\pm^{0.002}_{0.002}$ &  $\pm^{0.06}_{0.06}$ &   - &   - &   -  \\ \hline
    \multirow{2}{*}{2016-02 L$^\prime$ - 3PS} & $-60.5$ &  $0.083$ &  $4.42$ &  $-5.6$ &  $0.12$ &  $4.75$ &  $37.7$ &  $0.121$ &  $5.13$ \\
    & $\pm^{1.5}_{29.9}$ &  $\pm^{0.004}_{0.005}$ &  $\pm^{0.1}_{0.65}$ &  $\pm^{2.2}_{29.5}$ &  $\pm^{0.004}_{0.004}$ &  $\pm^{0.39}_{0.11}$ &  $\pm^{45.6}_{3.2}$ &  $\pm^{0.008}_{0.051}$ &  $\pm^{0.13}_{1.52}$ \\ \hline\hline
    \multirow{2}{*}{2016-11 L$^\prime$ - 1PS} & $8.9$ &  $0.089$ &  $4.94$ &   - &   - &   - &   - &   - &   -  \\
    & $\pm^{2.4}_{3.0}$ &  $\pm^{0.006}_{0.009}$ &  $\pm^{0.09}_{0.11}$ &   - &   - &   - &   - &   - &   - \\ \hline
    \multirow{2}{*}{2016-11 L$^\prime$ - 2PS} & $-58.0$ &  $0.091$ &  $4.73$ &  $7.2$ &  $0.08$ &  $4.57$ &   - &   - &   - \\
    & $\pm^{1.7}_{1.8}$ &  $\pm^{0.005}_{0.006}$ &  $\pm^{0.09}_{0.08}$ &  $\pm^{2.2}_{1.8}$ &  $\pm^{0.007}_{0.006}$ &  $\pm^{0.1}_{0.1}$ &   - &   - &   -  \\ \hline
    \multirow{2}{*}{2016-11 L$^\prime$ - 3PS} & $-56.9$ &  $0.096$ &  $4.68$ &  $-20.1$ &  $0.142$ &  $5.39$ &  $11.2$ &  $0.081$ &  $4.55$ \\
    & $\pm^{2.1}_{1.9}$ &  $\pm^{0.005}_{0.006}$ &  $\pm^{0.08}_{0.09}$ &  $\pm^{2.6}_{3.2}$ &  $\pm^{0.005}_{0.01}$ &  $\pm^{0.16}_{0.16}$ &  $\pm^{2.2}_{2.1}$ &  $\pm^{0.006}_{0.007}$ &  $\pm^{0.11}_{0.11}$ \\ \hline\hline
    \multirow{2}{*}{2018-01 L$^\prime$ - 1PS} & $-35.0$ &  $0.082$ &  $5.35$ &   - &   - &   - &   - &   - &   -  \\
    &  $\pm^{0.7}_{0.7}$ &  $\pm^{0.001}_{0.001}$ &  $\pm^{0.02}_{0.03}$ &   - &   - &   - &   - &   - &   - \\ \hline
    \multirow{2}{*}{2018-01 L$^\prime$ - 2PS} & $-51.6$ &  $0.094$ &  $5.26$ &  $-8.9$ &  $0.094$ &  $5.28$ &   - &   - &   -  \\
    & $\pm^{0.6}_{0.5}$ &  $\pm^{0.001}_{0.001}$ &  $\pm^{0.02}_{0.02}$ &  $\pm^{0.6}_{0.7}$ &  $\pm^{0.001}_{0.001}$ &  $\pm^{0.03}_{0.02}$ &   - &   - &   -  \\ \hline
    \multirow{2}{*}{2018-01 L$^\prime$ - 3PS} & $-84.4$ &  $0.102$ &  $5.97$ &  $-45.8$ &  $0.092$ &  $5.27$ &  $-3.5$ &  $0.096$ &  $5.33$   \\
    & $\pm^{2.0}_{2.1}$ &  $\pm^{0.001}_{0.001}$ &  $\pm^{0.06}_{0.06}$ &  $\pm^{1.6}_{1.4}$ &  $\pm^{0.001}_{0.001}$ &  $\pm^{0.02}_{0.02}$ &  $\pm^{0.9}_{0.9}$ &  $\pm^{0.001}_{0.001}$ &  $\pm^{0.04}_{0.03}$  \\ \hline\hline
    \multirow{2}{*}{2018-11 L$^\prime$ - 1PS} & $-33.8$ &  $0.084$ &  $5.4$ &   - &   - &   - &   - &   - &   -  \\
    & $\pm^{1.5}_{2.1}$ &  $\pm^{0.001}_{0.001}$ &  $\pm^{0.03}_{0.03}$ &   - &   - &   - &   - &   - &   - \\ \hline
    \multirow{2}{*}{2018-11 L$^\prime$ - 2PS} & $-59.5$ &  $0.093$ &  $5.36$ &  $-12.4$ &  $0.09$ &  $5.37$ &   - &   - &   -  \\
    & $\pm^{1.3}_{1.1}$ &  $\pm^{0.001}_{0.001}$ &  $\pm^{0.04}_{0.04}$ &  $\pm^{1.7}_{1.6}$ &  $\pm^{0.002}_{0.002}$ &  $\pm^{0.05}_{0.03}$ &   - &   - &   - \\ \hline
    \multirow{2}{*}{2018-11 L$^\prime$ - 3PS} & $-69.3$ &  $0.094$ &  $5.5$ &  $-29.1$ &  $0.096$ &  $5.33$ &  $11.6$ &  $0.092$ &  $5.69$ \\
    & $\pm^{1.5}_{1.3}$ &  $\pm^{0.002}_{0.001}$ &  $\pm^{0.05}_{0.05}$ &  $\pm^{2.2}_{1.9}$ &  $\pm^{0.001}_{0.001}$ &  $\pm^{0.05}_{0.04}$ &  $\pm^{2.4}_{2.7}$ &  $\pm^{0.002}_{0.001}$ &  $\pm^{0.06}_{0.06}$\\ \hline\hline
    \multirow{2}{*}{2019-01 H - 1PS} & $21.3$ &  $0.088$ &  $7.04$ &   - &   - &   - &   - &   - &   - \\
    & $\pm^{1.9}_{3.6}$ &  $\pm^{0.003}_{0.002}$ &  $\pm^{0.17}_{0.16}$ &   - &   - &   - &   - &   - &   - \\ \hline
    \multirow{2}{*}{2019-01 H - 2PS} & $-2.4$ &  $0.085$ &  $7.15$ &  $22.3$ &  $0.088$ &  $6.93$ &   - &   - &   -  \\
    & $\pm^{5.2}_{160.8}$ &  $\pm^{0.062}_{0.004}$ &  $\pm^{0.17}_{0.14}$ &  $\pm^{1.7}_{2.0}$ &  $\pm^{0.002}_{0.001}$ &  $\pm^{0.17}_{0.14}$ &   - &   - &   - \\ \hline
    \multirow{2}{*}{2019-01 H - 3PS} & $-55.5$ &  $0.076$ &  $7.29$ &  $4.6$ &  $0.083$ &  $7.07$ &  $24.8$ &  $0.085$ &  $7.04$  \\
    & $\pm^{14.3}_{108.0}$ &  $\pm^{0.071}_{0.045}$ &  $\pm^{0.21}_{0.17}$ &  $\pm^{18.3}_{7.5}$ &  $\pm^{0.005}_{0.007}$ &  $\pm^{0.21}_{0.2}$ &  $\pm^{11.5}_{4.3}$ &  $\pm^{0.004}_{0.043}$ &  $\pm^{0.24}_{0.23}$ \\ \hline\hline
    \multirow{2}{*}{2020-01 L$^\prime$ - 1PS} & $-39.0$ &  $0.086$ &  $5.59$ &   - &   - &   - &   - &   - &   -  \\
    & $\pm^{1.8}_{1.5}$ &  $\pm^{0.002}_{0.002}$ &  $\pm^{0.05}_{0.06}$ &   - &   - &   - &   - &   - &   -   \\ \hline
    \multirow{2}{*}{2020-01 L$^\prime$ - 2PS} & $-49.1$ &  $0.091$ &  $5.44$ &  $0.3$ &  $0.091$ &  $5.6$ &   - &   - &   - \\
    &  $\pm^{1.3}_{1.4}$ &  $\pm^{0.001}_{0.002}$ &  $\pm^{0.04}_{0.05}$ &  $\pm^{1.4}_{1.8}$ &  $\pm^{0.002}_{0.002}$ &  $\pm^{0.06}_{0.06}$ &   - &   - &   - \\ \hline
    \multirow{2}{*}{2020-01 L$^\prime$ - 3PS} & $-76.0$ &  $0.097$ &  $5.77$ &  $-37.1$ &  $0.094$ &  $5.4$ &  $8.3$ &  $0.097$ &  $5.64$ \\
    &  $\pm^{1.3}_{1.6}$ &  $\pm^{0.002}_{0.002}$ &  $\pm^{0.07}_{0.07}$ &  $\pm^{1.5}_{1.4}$ &  $\pm^{0.001}_{0.002}$ &  $\pm^{0.04}_{0.05}$ &  $\pm^{1.4}_{1.3}$ &  $\pm^{0.002}_{0.002}$ &  $\pm^{0.06}_{0.06}$ \\ \hline\hline
    \multirow{2}{*}{All L$^\prime$ - 1PS} & $-40.4$ &  $0.087$ &  $5.37$ &   - &   - &   - &   - &   - &   -  \\
    & $\pm^{0.5}_{0.5}$ &  $\pm^{0.001}_{0.001}$ &  $\pm^{0.02}_{0.02}$ &   - &   - &   - &   - &   - &   -  \\ \hline 
    \multirow{2}{*}{All L$^\prime$ - 2PS} & $-54.4$ &  $0.094$ &  $5.28$ &  $-9.1$ &  $0.092$ &  $5.37$ &   - &   - &   -  \\
    & $\pm^{0.5}_{0.5}$ &  $\pm^{0.001}_{0.001}$ &  $\pm^{0.02}_{0.01}$ &  $\pm^{0.5}_{0.7}$ &  $\pm^{0.001}_{0.001}$ &  $\pm^{0.02}_{0.02}$ &   - &   - &   -  \\ \hline 
    \multirow{2}{*}{All L$^\prime$ - 3PS}  & $-76.0$ &  $0.099$ &  $5.67$ &  $-37.6$ &  $0.094$ &  $5.33$ &  $3.4$ &  $0.096$ &  $5.53$ \\
    &  $\pm^{0.9}_{0.7}$ &  $\pm^{0.001}_{0.001}$ &  $\pm^{0.03}_{0.03}$ &  $\pm^{0.9}_{1.0}$ &  $\pm^{0.001}_{0.001}$ &  $\pm^{0.02}_{0.02}$ &  $\pm^{0.8}_{0.6}$ &  $\pm^{0.001}_{0.001}$ &  $\pm^{0.02}_{0.03}$  \\\hline
\enddata

\noindent \footnotesize{For each dataset and model type, the top row shows the best-fit parameter and the bottom row shows the uncertainties. Entries in each column are: (1) dataset and number of point sources (in addition to the central star) included in the model, with 1PS, 2PS, and 3PS indicating one, two, and three point sources, respectively; (2-4) position angle (measured in degrees east of north), separation (measured in arcseconds), and contrast (measured in magnitudes), respectively, of point source 1; (5-7) same as (2-4) for point source 2 (if applicable); (8-10) same as (2-4) for point source 3 (if applicable).}
\end{deluxetable*}

\begin{deluxetable*}{lccccccccccccc}
\tablecaption{Polar Gaussian Ring Fit Results\label{tab:diskfits}}
\tablewidth{700pt}
\tabletypesize{\scriptsize}
\tablehead{
\colhead{Fit} & \colhead{$r_i$} & \colhead{$\sigma_{r_i}$} & \colhead{$\theta_i$} & \colhead{$\sigma_{\theta_i}$} & \colhead{$r_{a_i}$} &  \colhead{$f_i$} & \colhead{$r_o$} & \colhead{$\sigma_{r_o}$} & \colhead{$\theta_o$} & \colhead{$\sigma_{\theta_o}$} & \colhead{$r_{a_o}$} & \colhead{$f_o$} & \colhead{$f_*$}\\ 
\colhead{(1)} & \colhead{(2)} & \colhead{(3)} & \colhead{(4)} & \colhead{(5)} & \colhead{(6)} &  \colhead{(7)} & \colhead{(8)} & \colhead{(9)} & \colhead{(10)} & \colhead{(11)} & \colhead{(12)} & \colhead{(13)} & \colhead{(14)}
} 
\startdata
   \multicolumn{14}{c}{Fits to All Observations in Each Band}\\ \hline\hline
   \multirow{2}{*}{All L$^\prime$} &   $124.4$ &  $23.1$ &  $330.6$ &  $45.0$ &  $0.68$ &  $0.025$ & $334.6$ & $37.9$ & $329.5$ & $117.1$ & $0.62$ & $0.042$ & $0.933$ \\
    & $\pm^{3.6}_{2.3}$ & $\pm^{1.9}_{1.8}$ & $\pm^{0.4}_{0.5}$ & $\pm^{1.4}_{1.2}$ & $\pm^{0.02}_{0.02}$ & $\pm^{0.002}_{0.003}$ & $\pm^{12.9}_{10.7}$ & $\pm^{8.8}_{7.0}$ & $\pm^{1.4}_{1.7}$ & $\pm^{5.7}_{5.4}$ & $\pm^{0.03}_{0.02}$ & $\pm^{0.001}_{0.001}$ & $\pm^{0.001}_{0.001}$ \\ \hline
    \multirow{2}{*}{2015-02 K$_\mathrm{s}$} & $143.4$ &  $25.0$ &  $322.7$ &  $42.5$ &  $0.62$ &  $0.031$ & $383.4$ & $10.8$ & $315.3$ & $35.6$ & $0.63$ & $0.02$ & $0.949$\\ 
    & $\pm^{24.9}_{8.2}$ & $\pm^{6.1}_{10.0}$ & $\pm^{3.2}_{2.8}$ & $\pm^{5.4}_{4.0}$ & $\pm^{0.07}_{0.13}$ & $\pm^{0.007}_{0.008}$ & $\pm^{9.4}_{9.3}$ & $\pm^{3.3}_{1.0}$ & $\pm^{20.3}_{1.5}$ & $\pm^{9.6}_{4.7}$ & $\pm^{0.03}_{0.11}$ & $\pm^{0.004}_{0.003}$ & $\pm^{0.004}_{0.003}$\\ \hline
    \multirow{2}{*}{2019-01 H} & $133.1$ &  $19.4$ &  $331.1$ &  $54.0$ &  $0.60$ &  $0.028$ & $351.8$ & $31.4$ & $328.1$ & $31.2$ & $0.55$ & $0.032$ & $0.940$ \\
    & $\pm^{4.4}_{4.0}$ & $\pm^{2.1}_{2.4}$ & $\pm^{1.5}_{1.7}$ & $\pm^{6.8}_{4.8}$ & $\pm^{0.03}_{0.02}$ & $\pm^{0.009}_{0.01}$ & $\pm^{16.0}_{15.8}$ & $\pm^{3.0}_{2.5}$ & $\pm^{1.2}_{0.8}$ & $\pm^{2.0}_{1.5}$ & $\pm^{0.03}_{0.02}$ & $\pm^{0.004}_{0.003}$ & $\pm^{0.006}_{0.006}$ \\ \hline\hline
    \multicolumn{14}{c}{Fits to Individual L$^\prime$ Epochs}\\ \hline\hline
    \multirow{2}{*}{2014-12 L$^\prime$} & $118.9$ &  $32.5$ &  $325.9$ &  $73.8$ &  $0.53$ &  $0.055$ & $344.1$ & $129.8$ & $318.9$ & $11.1$ & $0.54$ & $0.012$ & $0.932$  \\
    & $\pm^{15.4}_{24.2}$ & $\pm^{21.1}_{12.2}$ & $\pm^{3.2}_{3.5}$ & $\pm^{10.1}_{13.3}$ & $\pm^{0.07}_{0.07}$ & $\pm^{0.009}_{0.015}$ & $\pm^{40.1}_{54.9}$ & $\pm^{52.9}_{75.7}$ & $\pm^{6.8}_{3.4}$ & $\pm^{69.1}_{4.1}$ & $\pm^{0.09}_{0.07}$ & $\pm^{0.01}_{0.004}$ & $\pm^{0.005}_{0.006}$  \\ \hline
    \multirow{2}{*}{2016-02 L$^\prime$} & $110.9$ &  $24.2$ &  $351.8$ &  $61.4$ &  $0.41$ &  $0.075$ & $392.2$ & $18.4$ & $347.9$ & $22.5$ & $0.44$ & $0.019$ & $0.906$\\
    & $\pm^{10.9}_{6.4}$ & $\pm^{17.5}_{7.0}$ & $\pm^{2.3}_{8.6}$ & $\pm^{5.2}_{3.8}$ & $\pm^{0.34}_{0.09}$ & $\pm^{0.008}_{0.038}$ & $\pm^{6.6}_{55.2}$ & $\pm^{12.6}_{1.9}$ & $\pm^{3.8}_{4.3}$ & $\pm^{32.2}_{3.8}$ & $\pm^{0.34}_{0.1}$ & $\pm^{0.033}_{0.004}$ & $\pm^{0.005}_{0.004}$\\ \hline
    \multirow{2}{*}{2016-11 L$^\prime$} & $64.8$ &  $45.3$ &  $340.4$ &  $74.3$ &  $0.79$ &  $0.08$ & $307.4$ & $18.2$ & $337.4$ & $76.8$ & $0.78$ & $0.061$ & $0.858$  \\
    & $\pm^{33.0}_{27.3}$ & $\pm^{12.8}_{16.1}$ & $\pm^{3.3}_{2.7}$ & $\pm^{19.1}_{16.8}$ & $\pm^{0.04}_{0.06}$ & $\pm^{0.075}_{0.065}$ & $\pm^{26.5}_{14.3}$ & $\pm^{2.7}_{1.4}$ & $\pm^{3.0}_{4.0}$ & $\pm^{27.1}_{19.3}$ & $\pm^{0.04}_{0.06}$ & $\pm^{0.035}_{0.021}$ & $\pm^{0.03}_{0.054}$  \\ \hline
    \multirow{2}{*}{2018-01 L$^\prime$} & $115.4$ &  $19.3$ &  $327.5$ &  $41.7$ &  $0.74$ &  $0.023$ & $275.1$ & $45.0$ & $326.2$ & $93.9$ & $0.72$ & $0.045$ & $0.932$ \\
    & $\pm^{3.1}_{2.3}$ & $\pm^{3.6}_{1.8}$ & $\pm^{0.9}_{1.0}$ & $\pm^{2.1}_{1.6}$ & $\pm^{0.02}_{0.02}$ & $\pm^{0.003}_{0.003}$ & $\pm^{9.0}_{9.5}$ & $\pm^{6.0}_{6.3}$ & $\pm^{3.6}_{1.5}$ & $\pm^{4.2}_{4.3}$ & $\pm^{0.04}_{0.03}$ & $\pm^{0.001}_{0.002}$ & $\pm^{0.002}_{0.001}$ \\ \hline
    \multirow{2}{*}{2018-11 L$^\prime$} & $135.8$ &  $44.0$ &  $329.9$ &  $48.6$ &  $0.56$ &  $0.034$ & $391.2$ & $41.7$ & $331.6$ & $76.8$ & $0.53$ & $0.032$ & $0.934$ \\
    & $\pm^{7.5}_{4.6}$ & $\pm^{4.1}_{4.4}$ & $\pm^{1.0}_{1.0}$ & $\pm^{6.1}_{3.8}$ & $\pm^{0.02}_{0.03}$ & $\pm^{0.007}_{0.005}$ & $\pm^{6.0}_{8.9}$ & $\pm^{9.6}_{8.1}$ & $\pm^{1.8}_{1.7}$ & $\pm^{10.6}_{9.7}$ & $\pm^{0.02}_{0.01}$ & $\pm^{0.004}_{0.005}$ & $\pm^{0.001}_{0.002}$ \\ \hline
    \multirow{2}{*}{2020-01 L$^\prime$} & $144.5$ &  $38.1$ &  $330.8$ &  $50.2$ &  $0.55$ &  $0.034$ & $389.0$ & $63.3$ & $335.0$ & $50.0$ & $0.54$ & $0.022$ & $0.944$\\
    & $\pm^{8.0}_{7.6}$ & $\pm^{5.3}_{4.0}$ & $\pm^{1.0}_{1.2}$ & $\pm^{5.0}_{4.2}$ & $\pm^{0.03}_{0.04}$ & $\pm^{0.007}_{0.009}$ & $\pm^{7.7}_{15.2}$ & $\pm^{17.9}_{13.8}$ & $\pm^{2.8}_{3.2}$ & $\pm^{13.7}_{8.0}$ & $\pm^{0.03}_{0.02}$ & $\pm^{0.005}_{0.004}$ & $\pm^{0.004}_{0.003}$ \\ \hline
\enddata

\noindent \footnotesize{For each dataset, the top row shows the best-fit parameter and the bottom row shows the uncertainties. Entries in each column are: (1) dataset used for model fitting; (2) inner ring radius in milliarcseconds; (3) inner ring radial standard deviation in milliarcseconds; (4) inner ring peak flux position angle in degrees; (5) inner ring azimuthal standard deviation in degrees; (6) inner ring axis ratio; (7) inner ring fractional flux; (8) outer ring radius in milliarcseconds; (9) outer ring radial standard deviation in milliarcseconds; (10) outer ring peak flux position angle in degrees; (11) outer ring azimuthal standard deviation in degrees; (12) outer ring axis ratio; (13) outer ring fractional flux; (14) central unresolved component fractional flux.}
\end{deluxetable*}

\subsection{Geometric Models}\label{sec:geomres} 
Tables \ref{tab:compfits} and \ref{tab:diskfits} list the best fit parameters and uncertainties for the multiple point source and polar Gaussian ring models, respectively.
Table \ref{tab:modelsel} lists goodness-of-fit metrics, including Bayesian evidence, $\chi^2$, and reduced $\chi^2$ values for each model type, as well as the null model (which is a single, unresolved star). 
Comparison of the $\chi^2$ values for all model types to the null model $\chi^2$ values shows that both the multiple point source and the polar Gaussian ring models are strongly preferred over the null model.
In the subsections that follow we describe the fit results and discuss whether the data prefer a static or dynamic scenario within each geometric model class, before comparing the two classes of models.  

\subsubsection{Multiple Point Source Fit Results}\label{sec:compres}
Table \ref{tab:compfits} shows the results of one-, two-, and three-point-source model fits to the individual epochs and to the combined L$^\prime$ data. 
The best fits generally place the three sources from the northeast to the west of the star, spanning position angles from $\sim30^\circ$ to $-90^\circ$ (along the same arcs seen in the reconstructed images). 
The separations range from $\sim70$ to $120$ mas, corresponding to 11-20 AU given the distance to LkCa 15. 
Best-fit contrasts at L$^\prime$ band are consistent with similar geometric model results from previous studies, at $\sim5-6$ magnitudes \citep[e.g.][]{2012ApJ...745....5K,2015Natur.527..342S}.

Given that the one-, two-, and three-point-source models are nested, it is straightforward to use both Bayesian evidence \citep[e.g.][]{2008ConPh..49...71T,2011MNRAS.413.2895J} and $\chi^2$ intervals to perform model selection without complications related to priors on parameters (Table \ref{tab:modelsel}). 
Each additional point source adds three model parameters. 
The best-fit $\chi^2$ values can thus be compared to a distribution with three degrees of freedom, where a 5$\sigma$ model preference corresponds to a $\chi^2$ improvement ($\Delta \chi^2$) of 31.81.
For all observational epochs, going from a one- to two-point-source model and going from a two- to three-point source model results in a larger $\chi^2$ improvement than this.
These $\chi^2$ improvements are in agreement with the log evidence values, which increase significantly as each source is added and by $\sim150-500$ with the addition of the third.
The three-point-source model is thus significantly preferred over the one- and two-point-source models. 

For most cases, the improvement in $\chi^2$ is $\gtrsim100-300$ for the addition of the second and third sources, suggesting that even if the error bars were under-estimated the three-point-source model would still be selected over the others. 
The least significant improvements are going from the two- to three-point-source model for the 2014-12 L$^\prime$ band, 2016-11 L$^\prime$ band, and 2019-01 H band datasets, which range from $\Delta \chi^2$ = 35-47.
Here, under-estimated error bars by a factor of $\sim2$ (a scenario potentially supported by the high reduced $\chi^2$ values) would reduce the significance of the model preference from 5$\sigma$ to $2-3\sigma$. 
Increasing the error bars would decrease the Bayesian evidence values uniformly (since evidence is likelihood marginalized over the parameter space). 
However, since log evidence improvements are taken to be significant when they are greater than $\sim5$ \citep[e.g.][]{2008ConPh..49...71T}, the log evidence improvements would still be significant with scaled-up error bars (at $\sim37-125$ going from two to three point sources).

Assuming the three-point-source model is preferred (within this model class), we can use the parameters in Table \ref{tab:compfits} and the $\chi^2$ values in Table \ref{tab:modelsel} to determine whether the data support a dynamic or static multiple-point-source scenario. 
The top half of Table \ref{tab:idisk_dyntest} shows the differences between the best-fit $\chi^2$ values for each epoch and the $\chi^2$ value of the static scenario parameters at that epoch. 
We calculate these differences two ways: first without any re-scaling of the $\chi^2$ values, and second by calculating the $\chi^2$ interval after scaling so that the best-fit reduced $\chi^2$ is equal to 1. 
Taking the conservative estimates, the dynamic scenario is preferred at at least 3$\sigma$ for all of the individual epochs, with a significance of $>5\sigma$ for the 2014-12 L$^\prime$, 2016-02 L$^\prime$, 2016-11 L$^\prime$, and 2018-01 L$^\prime$.
The 2018-11 L$^\prime$ and 2020-01 L$^\prime$ epochs prefer the dynamic scenario with 4$\sigma$ and 3$\sigma$ significance, respectively. 

We can also compare the $\chi^2$ values for the static and dynamic three-point-source fits to the combined multi-epoch L$^\prime$ dataset. 
The static scenario has nine free parameters, while the dynamic one has 54 (nine for each of the six L$^\prime$ epochs). 
Performing model selection by comparing the $\chi^2$ improvement (752.24) to a distribution for 45 degrees of freedom shows that the dynamic scenario is preferred at greater than 5$\sigma$ significance with or without rescaling to make the best-fit reduced $\chi^2$ equal to one.
Within the framework of the multiple point source model, fits to the individual epochs and to the combined L$^\prime$ datasets strongly support a dynamic scenario over a static one. 

We can quantify the significance of the variation in best-fit point source parameters by comparing each epoch's estimates to the best-fit static three-point-source model. 
The results are shown in the top half of Table \ref{tab:idisk_dyntest}, which focuses on quantifying the significance of position angle evolution. 
The majority of the best-fit position angles (15 of 18) differ significantly from the static model estimates. 
Of these, eight are $\geq3\sigma$ discrepancies and seven are $1-2\sigma$.
These statistical tests show that the multiple point source models prefer a dynamic scenario with significant positional evolution from epoch to epoch.

\subsubsection{Polar Gaussian Ring Fit Results}\label{sec:diskres}
Table \ref{tab:diskfits} lists the results of polar Gaussian ring fitting to the individual epochs and wavelengths. 
Here and for the remainder of the paper we focus on the two ring model results because all epochs strongly preferred the two ring model to the single ring model  (at $4-5\sigma$ based on improvements in best-fit $\chi^2$).
Figure \ref{fig:2disk_bands} shows best fit two ring models for combined observations in each band, and Figure \ref{fig:2disk_epochs} shows the best fit models for the individual L$^\prime$ epochs.
As shown in Figures \ref{fig:2disk_bands} and \ref{fig:2disk_epochs} and Table \ref{tab:diskfits}, the best-fit models consists of an inner ring with a radius of approximately 110-140 milliarcseconds (mas) and a Gaussian radial cross section with a standard deviation of 20-40 mas.
These scales correspond to 18-22 AU and 3-6 AU, respectively, at the distance of LkCa 15 (159 pc).
The outer ring radius ranges from 275-390 mas with a standard deviation of 11-60 mas, depending on the epoch and wavelength. 
The orientations for all epochs are such that a bright arc exists to the northwest of the star.

We use the parameters listed in Table \ref{tab:diskfits} and the $\chi^2$ values in Table \ref{tab:modelsel} to assess whether the multi-epoch observations are better explained by a static or dynamic scenario (within the parameterization of the polar Gaussian ring model).
We take the difference between the best-fit $\chi^2$ values for each epoch and the $\chi^2$ value of the static parameters at that epoch (Table \ref{tab:idisk_dyntest}, bottom half).
As for the multiple-point-source fits, we do this first without any re-scaling of the $\chi^2$ values, and second by calculating the $\chi^2$ interval after scaling so that the best-fit reduced $\chi^2$ is equal to 1. 
In both cases, all epochs show at least a 3$\sigma$ preference for the dynamic model over the static model. 
Taking the more conservative estimate for each epoch, the 2016-02 L$^\prime$, 2018-01 L$^\prime$, 2018-11 L$^\prime$, and 2020-01 L$^\prime$ datasets prefer the dynamic model at $>5\sigma$ significance; the 2016-11 L$^\prime$ at $>4\sigma$ significance, and the 2014-12 L$^\prime$ at $>3\sigma$ significance.

We also compare the $\chi^2$ value of the All L$^\prime$ static fit to its dynamic fit $\chi^2$ (which is the summed $\chi^2$ values of the polar Gaussian ring fits to all of the individual L$^\prime$ epochs; Table \ref{tab:modelsel}, bottom half). 
The static fit has 12 free parameters, while the dynamic fit has 72 (the 12 polar Gaussian ring parameters repeated for each of the six L$^\prime$ epochs). 
Comparing the improvement in $\chi^2$ (1013.51) to a distribution with 60 degrees of freedom (the difference in free parameters between the two models) shows that the dynamic model is preferred with $>5\sigma$ significance.
This is true even if the error bars are scaled up so that the reduced $\chi^2$ of the dynamic model is equal to 1. 
Under the assumption of the polar Gaussian ring model class, both the individual epochs and the combined L$^\prime$ dataset thus strongly prefer a dynamic scenario to a static one. 

We also search for variability in the best-fit polar Gaussian ring parameters by comparing each epoch's best fit to the static fit parameters. 
We focus on the significance of discrepancies between the dynamic and static inner ring parameters since the inner ring component lies at the separation of the companion candidates. 
In agreement with the evolving geometry in Figure \ref{fig:2disk_epochs}, the majority of the best-fit inner ring parameters for the individual epochs (24 of 36; Table \ref{tab:idisk_dyntest}) disagree with the static model at at least the 1$\sigma$ level.
Of these, 16 are 1$\sigma$ discrepancies, five are 2$\sigma$, and three are 3$\sigma$. 
Each epoch has at least three discrepant parameters (of six total) compared to the static model.
Furthermore, all of the six inner ring parameters differ from the static estimates in three to four of the six epochs, suggesting complex variations.
These two statistical tests show that - under the assumption of the polar Gaussian ring model - the dynamic scenario is strongly preferred and the source morphology evolves significantly and in a complex way.

\begin{deluxetable*}{lccccccc}
\tablecaption{Goodness of Fit Metrics\label{tab:modelsel}}
\tablewidth{700pt}
\tabletypesize{\scriptsize}
\tablehead{
\colhead{Dataset} & \colhead{Null Model} & \colhead{2PG rings: static$^*$} & \colhead{2PG rings: dynamic} & \colhead{1PS: dynamic} & \colhead{2PS: dynamic} & \colhead{3PS: dynamic} & \colhead{3PS: static$^*$}
}
\startdata
\multicolumn{8}{c}{Bayesian Evidence} \\ \hline\hline
 2014-12 L$^\prime$ & -- & -- & $-630.34\pm0.02$& $-4932.346\pm0.017$ & $-578.90\pm0.016$& $-438.77\pm0.04$ & -- \\
 2015-02 K$_\mathrm{s}$ & -- & $-2221.637\pm0.002$ & $-2221.637\pm0.002$ & $-7304.28\pm0.03$ & $-1628.244\pm0.006$& $-1463.921\pm 0.004$ & $-1463.921\pm 0.004$ \\
 2016-02 L$^\prime$ & -- & -- &$-2501.12\pm0.08$ & $-8190.436\pm0.001$& $-2560.658\pm0.005$& $-2324.48\pm0.01$ & --\\
 2016-11 L$^\prime$ & -- & -- & $-628.40\pm0.019$ & $-6874.17\pm0.05$& $-820.91\pm0.01$& $-683.87\pm0.03$ & --\\
 2018-01 L$^\prime$ & -- & -- & $-2673.34\pm98.59$& $-9612.969\pm0.013$ & $-2649.326\pm0.006$& $-2145.58\pm0.03$ & --\\
 2018-11 L$^\prime$ & -- & -- & $-1118.89\pm35.39$ & $-8838.295\pm0.008$ & $-1283.61\pm0.02$ & $-945.98\pm0.22$ & -- \\
 2019-01 H & -- &  $-2280.69\pm3.32$ & $-2280.69\pm3.32$& $-7393.212\pm0.002$& $-2385.81\pm0.05$ & $-2248.20\pm0.06$ & $-2248.20\pm0.06$ \\
 2020-01 L$^\prime$ & -- & -- & $-1016.97\pm2.12$ & $-7994.177\pm 0.008$ & $-1192.384\pm0.007$ & $-955.21\pm 0.07$ & -- \\  \hline\hline
 \multicolumn{8}{c}{Chi-Squared$^\dagger$} \\ \hline\hline
 2014-12 L$^\prime$ & 1464.88 (1.49) & 993.54 (1.02) & 957.31 (0.99) & 731.56 (0.75) & 640.44 (0.66) & 605.66 (0.62) & 672.68 (0.69) \\
 2015-02 K$_\mathrm{s}$  & 4465.26 (3.99) & 3961.07 (3.71) & 3961.07 (3.57) & 2756.17 (2.47) & 2635.63 (2.37) & 2537.41 (2.28) & 2537.41 (2.28)  \\
 2016-02 L$^\prime$  & 6019.64 (6.61) & 4645.23 (5.17) & 4222.57 (4.71) & 4473.27 (4.93) & 4135.82 (4.58) & 3930.80 (4.36) & 4338.70 (4.82) \\
 2016-11 L$^\prime$  & 1398.24 (3.33) & 1124.60 (2.76) & 1013.63 (2.48) & 1249.50 (3.00) & 1074.90 (2.60) & 1028.80 (2.50) & 1147.37 (2.79) \\
 2018-01 L$^\prime$  & 10320.98 (4.53) & 3593.80 (1.58) & 3486.18 (1.54) & 4822.34 (2.12) & 3269.56 (1.44) & 2882.70 (1.27) & 2959.24 (1.30) \\
 2018-11 L$^\prime$  & 4878.11 (3.70) & 854.40 (0.65) & 783.99 (0.60) & 2124.84 (1.61) & 1262.01 (0.96) & 936.75 (0.71) & 978.40 (0.75) \\
 2019-01 H  & 4481.57 (1.44) & 4182.54 (1.35) & 4182.54 (1.35) & 4290.28 (1.38) & 4238.72 (1.36) & 4194.98 (1.35) & 4194.98 (1.35)\\
 2020-01 L$^\prime$  & 2750.42 (1.43) & 1346.57 (0.71) & 1291.89 (0.68) & 1863.96 (0.97) & 1430.90 (0.75) & 1217.50 (0.64) & 1249.78 (0.65) \\
  All L$^{\prime\ddagger}$  & 26832.27 (3.43) & 12769.08 (1.63)  & 11755.57 (1.52) & 15265.47 (1.95) & 11813.63 (1.52) & 10602.21 (1.36) & 11354.45 (1.45) \\ 
\enddata
\footnotesize{$^*$ Since Bayesian evidence is marginalized over the parameter space, we cannot calculate values for the static polar Gaussian ring model or static three-point-source model for the individual epochs. The best-fit parameters for these static scenarios represent just one point in the explored parameter space for each L$^\prime$ epoch. Since only one $\mathrm{K_s}$ and H band epoch are considered, the static and dynamic models are identical for those datasets.}

\footnotesize{$^\dagger$ Each column in the Chi-Squared section of the table shows the raw $\chi^2$ followed by the reduced $\chi^2$ in parentheses.}

\footnotesize{$^\ddagger$ We only calculate $\chi^2$ values for the static and dynamic fits to the combined L$^\prime$ datasets. This is because fully exploring the parameter space for the dynamic models (which have a large number of free parameters) to calculate the Bayesian evidence would be extremely computationally expensive.}
\end{deluxetable*}

\begin{deluxetable*}{lcccccc}
\tablecaption{Geometric Model Selection: Static Versus Dynamic Scenarios$^*$ \label{tab:idisk_dyntest}}
\tablewidth{700pt}
\tabletypesize{\scriptsize}
\tablehead{
\colhead{Parameter} & \colhead{2014-12 L$^\prime$} & \colhead{2016-02 L$^\prime$} &\colhead{2016-11 L$^\prime$} & \colhead{2018-01 L$^\prime$} & \colhead{2018-11 L$^\prime$} & \colhead{2020-01 L$^\prime$}
} 
\startdata\multicolumn{7}{c}{Three Point Source Models: Dynamic Scenario Model Preference} \\ \hline
Significance ($\Delta \chi^2$)$^\dagger$ & $>5\sigma$ (67.02) & $>5\sigma$ (408.1) & $>5\sigma$ (118.57) & $>5\sigma$ (76.54) & $>4\sigma$ (41.65) & $>3\sigma$ (32.28)\\
Significance (Renormed $\Delta \chi^2$)$^\ddagger$ & $>5\sigma$ (108.10) & $>5\sigma$ (93.6) & $>5\sigma$ (47.43) & $>5\sigma$ (60.27) & $>5\sigma$ (58.66) & $>5\sigma$ (49.66) \\\hline
\multicolumn{7}{c}{Three Point Source Models: Position Angle Discrepancies w.r.t Static Best Fit} \\ \hline
$\mathrm{PA_1}$ (Point Source 1 Position Angle) & 2$\sigma$ & -- & $>5\sigma$ & 3$\sigma$ & 3$\sigma$ & -- \\
$\mathrm{PA_2}$ (Point Source 2 Position Angle) & 1$\sigma$ & 1$\sigma$ & 4$\sigma$ & 2$\sigma$ & 3$\sigma$ & --\\
$\mathrm{PA_3}$ (Point Source 3 Position Angle) & 3$\sigma$ & $>5\sigma$ & 2$\sigma$ & 4$\sigma$ & 2$\sigma$ & 2$\sigma$\\ \hline \hline
\multicolumn{7}{c}{Polar Gaussian Ring Models: Dynamic Scenario Model Preference} \\ \hline
Significance ($\Delta \chi^2$)$^\dagger$ & $>3\sigma$ (36.03) & $>5\sigma$ (422.66)& $>5\sigma$ (110.97) & $>5\sigma$ (107.62) & $>5\sigma$ 
 (70.41) & $>5\sigma$ (54.68) \\
Significance (Renormed $\Delta \chi^2$)$^{\ddagger}$ & $>3\sigma$ (36.39) & $>5\sigma$ (89.74)& $>4\sigma$ (44.74) & $>5\sigma$ (69.88) & $>5\sigma$ (117.35) & $>5\sigma$ (80.41) \\ \hline
\multicolumn{7}{c}{Polar Gaussian Ring Models: Inner Ring Parameter Discrepancies w.r.t Static Best Fit} \\ \hline
$r_i$ (Radius) & -- & -- & 1$\sigma$ & 1$\sigma$ & 1$\sigma$ & 1$\sigma$\\
$\sigma_{r_i}$ (Radial Stddev.) & -- & -- & 1$\sigma$ & -- & 3$\sigma$ & 2$\sigma$\\
$\theta_i$ (Position Angle) & 1$\sigma$ & 2$\sigma$ & 3$\sigma$ & 2$\sigma$ & -- & --\\
$\sigma_{\theta_i}$ (Azimuthal Stddev.) & 1$\sigma$ & 3$\sigma$ & 1$\sigma$ & 1$\sigma$ & -- & --\\
$r_{a_i}$ (Axis Ratio) & 1$\sigma$ & -- & 1$\sigma$ & 1$\sigma$ & 2$\sigma$ & 2$\sigma$\\
$f_i$ (Fractional Flux) & 1$\sigma$ & 1$\sigma$ & -- & -- & 1$\sigma$ &  1$\sigma$ 
\enddata
\footnotesize{$^*$ This table compares the static and dynamic scenarios for each geometric model class, with the multiple-point-source model above the double line and the polar Gaussian ring model below the double line. For each model class, the first two rows show the significance with which the dynamic scenario is preferred for each individual epoch of observations. The remaining rows show discrepancies between relevant parameters estimated for each L$^\prime$ epoch and those estimated for a static fit to all L$^\prime$ epochs.}

\footnotesize{$^\dagger$ Significance of the preference for the dynamic model. Calculated from the $\chi^2$ difference between each epoch's best fit and the static model (shown in parentheses), using a $\chi^2$ distribution with 12 degrees of freedom for the polar Gaussian ring model and 9 degrees of freedom for the three-point-source model.}

\footnotesize{$^\ddagger$ Significance of the preference for the dynamic model.  Calculated from the $\chi^2$ difference between each epoch's best fit and the static model (shown in parentheses) after rescaling so that the best-fit reduced $\chi^2$ is equal to 1, using a $\chi^2$ distribution with 12 degrees of freedom for the polar Gaussian ring model and 9 degrees of freedom for the three-point-source model.}
\end{deluxetable*}

\subsubsection{Goodness-of-Fit Tests: Multiple Point Source Versus Polar Gaussian Ring Models}\label{sec:geomfitgoodness}

We can compare the goodness-of-fit metrics for the multiple point source and polar Gaussian ring models to test whether the data prefer one set of models over the other.
This can allow us to distinguish between a ``clumpy" or smooth brightness distribution. 
The $\chi^2$ metrics presented in Table \ref{tab:modelsel} show that when all L$^\prime$ epochs are combined, the three-point-source model provides a better fit than the polar Gaussian ring model, suggesting a clumpier morphology. 
Examining the All L$^\prime$ values for the static and dynamic model types shows that both the static and dynamic three-point-source models are preferred over the polar Gaussian ring models, with the following order in the model preferences: (1) dynamic three-point-source, (2) static three-point-source, (3) dynamic polar Gaussian ring, (4) static polar Gaussian ring.
The single-epoch $\mathrm{K_s}$ data strongly prefer the three-point-source model to the polar Gaussian ring model, while the single-epoch H band data marginally prefer the polar Gaussian ring model to the three-point-source model. 

For the individual L$^\prime$ epochs, all datasets prefer the dynamic three-point-source model, except the 2016-11 L$^\prime$ and the 2018-11 L$^\prime$, which prefer the dynamic polar Gaussian ring.
Of these two, the 2016-11 L$^\prime$ epoch prefers the dynamic three-point-source model to the static polar Gaussian ring.
The 2018-11 L$^\prime$ epoch prefers the static polar Gaussian ring to the dynamic three-point-source model. 

In general, the reduced $\chi^2$ values for both model classes vary from $\sim0.6-5$ depending on the dataset.
For example, for the 2018-11 L$^\prime$ epoch the reduced $\chi^2$ values imply that the polar Gaussian ring model leads to worse over-fitting than the multiple point source model, while the opposite is true for other epochs such as 2014-12 L$^\prime$. 
This may result from uncertainty in systematic errors (under the assumption that each model is similarly adequate from epoch to epoch).
It may also indicate varying levels of complexity from epoch to epoch (such that the simple geometric models capture the true source morphology to varying degrees).
This type of variation would be consistent with the individual model classes' preferences for dynamic over static scenarios.

\citet{2022ApJ...931....3B} used Bayesian evidence values as model selectors for polar Gaussian ring versus multiple point source fitting to K band SPHERE NRM data, with the ring model maximizing the evidence. 
Examination of the evidence values in Table \ref{tab:modelsel} shows that, contrary to the results in \citet{2022ApJ...931....3B}, the three-point-source model has higher Bayesian evidence than the ring models for all epochs except 2016-11 L$^\prime$. 
While this suggests that the three-point-source model is preferred, Bayesian evidence should be used with caution when comparing non-nested models with different parameter priors.
Since the evidence is a marginal likelihood integrated over allowed parameter space, having more- or less-restrictive priors across models can bias their respective evidence values \citep[e.g.][]{2011MNRAS.413.2895J}. 
For example, the evidence improvement from the polar Gaussian ring to three-point-source model could be caused by the polar Gaussian ring model priors allowing it to explore larger regions of lower-likelihood parameter space. 
Despite these uncertainties, the $\chi^2$ preference for the three-point-source model over the polar Gaussian ring model (for most datasets) suggests that LkCa 15 has a clumpy morphology not easily captured by the polar Gaussian ring model prescription. 

\subsection{Image Reconstructions of Geometric Models}\label{sec:imreconsims}
We use the image reconstruction simulations (Section \ref{sec:imrecon_mod}) to explore the quality of the various geometric model fits as well. 
We simulate reconstructed images for the highest-quality individual L$^\prime$ epochs, using static and dynamic model types from the two classes as inputs.
By comparing all four of these models we can assess which source morphology is most consistent with the observed reconstructions, and whether the reconstructions show a preference for the dynamic or static scenarios for each model type.
We also reconstruct static polar Gaussian ring and static three-point-source models for the single H and $\mathrm{K_s}$ datasets, as well as the combined L$^\prime$ data.

\begin{figure*}
\begin{center}
\includegraphics[width=0.75\textwidth]{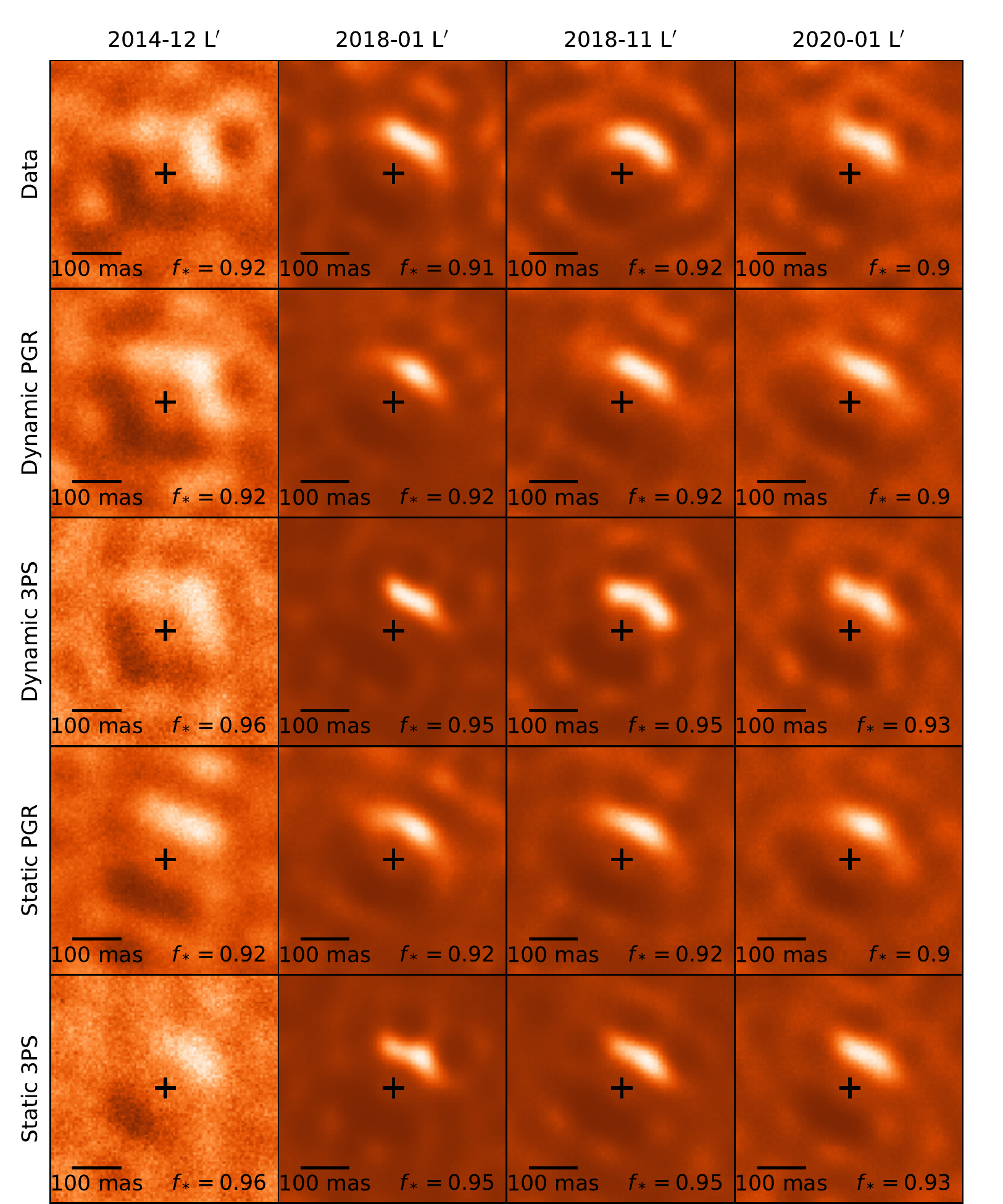}
\end{center}
\caption{Comparison of observed multi-epoch L$^\prime$ reconstructions to expectations for best-fit three-point-source and polar Gaussian ring models (both static and dynamic). The top row shows images reconstructed from the the highest signal-to-noise L$^\prime$ observations (orange; a subset of those shown in Figure \ref{fig:recons_epochs}). The following rows show images reconstructed from the best-fit dynamic polar Gaussian ring (Dynamic PGR), dynamic three-point-source (Dynamic 3PS), static polar Gaussian ring (Static PGR), and static three-point-source (Static 3PS), with Gaussian noise added to match the scatter in the observations. \label{fig:L_disk_comp_recons}}
\end{figure*}

\begin{figure}
\begin{center}
\includegraphics[width=\columnwidth]{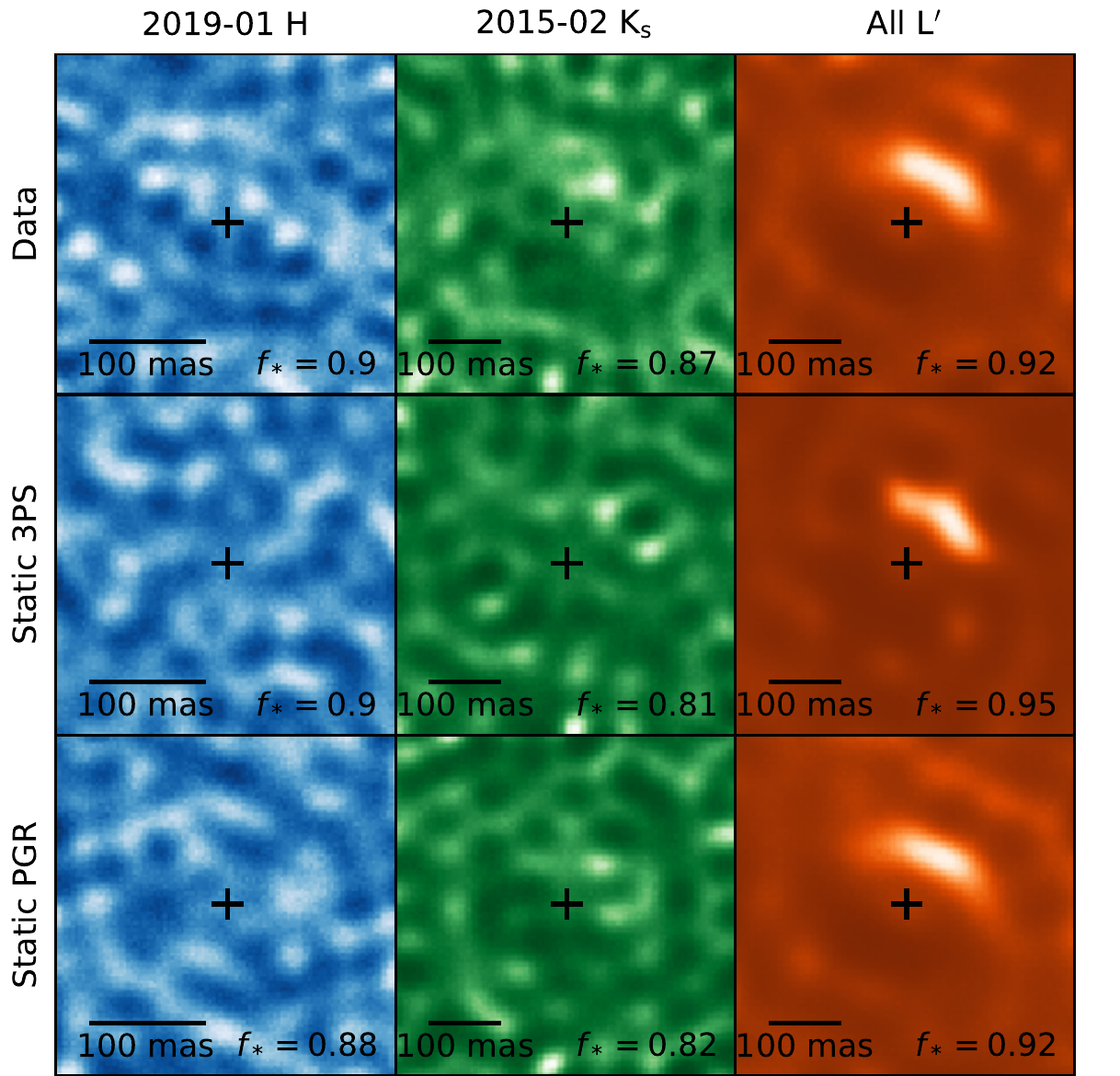}
\end{center}
\caption{Comparison of observed reconstructed images in combined datasets from each band (top row) to expectations for best-fit three-point-source (middle row) and polar Gaussian ring models (bottom row). From left to right columns show the 2019-01 H band, 2015-02 $\mathrm{K_s}$ band, and 2014-2020 combined L$^\prime$ band data. As fits to the combined multi-epoch L$^\prime$ dataset, the geometric models used here are the static models from Figure \ref{fig:L_disk_comp_recons}. Similar to Figure \ref{fig:L_disk_comp_recons}, we add Gaussian noise to each simulated observation to match the scatter in the real data.
The simulated reconstructions shown here are affected by both the Fourier coverage and noise levels of the mock observations. See Appendix \ref{sec:app_recons} for example simulated reconstructions where these effects are isolated.
\label{fig:comb_disk_comp_recons}}
\end{figure}

Figure \ref{fig:L_disk_comp_recons} shows the results for the individual L$^\prime$ epochs, and Figure \ref{fig:comb_disk_comp_recons} shows the results for the combined datasets at each bandpass.
For the individual L$^\prime$ epochs, the three-point-source models (both static and dynamic) better match the broad position angle extent of the emission and the sharp edges of the arc to the northwest of the star.  
This illustrates the limitations of the polar Gaussian model, which has a shallower flux drop-off from the peak position angle. 
Comparing the polar Gaussian ring and three-point-source reconstructions for each epoch suggests that the source brightness distribution is clumpier than the polar-Gaussian model allows, in agreement with the goodness-of-fit tests in Section \ref{sec:geomfitgoodness}.

Comparing the static and dynamic models in Figure \ref{fig:L_disk_comp_recons} shows that the reconstructions also prefer the dynamic scenarios.
While the reconstructions from each epoch exhibit the same general morphology of an arc to the northwest of the star, the extent and peak position angle of the arc shifts slightly from epoch to epoch.
The static models (both polar Gaussian ring and three-point-source) do not exhibit such changes at the same level. 
The different noise levels and amounts of sky rotation in each epoch result in small changes in the simulated static model reconstructions, but the extent and orientation of the arc are much more consistent than both the data and the dynamic model reconstructions. 
This is in agreement with the geometric fit preferences for the dynamic model types described in Section \ref{sec:diskres}.

For the combined L$^\prime$ data in Figure \ref{fig:comb_disk_comp_recons}, the arc in the observed reconstruction has edges that are not as sharp as the three-point-source fit, but are sharper than the polar Gaussian ring fit.
For this dataset, the three-point-source model still provides a better match than the polar Gaussian ring model, as evidenced by the lower $\chi^2$ value in Table \ref{tab:modelsel}.
However, the fact that the three-point-source model offers more improvement over the polar Gaussian ring model for the individual reconstructions may suggest that the combined L$^\prime$ reconstruction washes out variability between the individual epochs.

Lastly, as shown in Figure \ref{fig:comb_disk_comp_recons}, neither the polar Gaussian ring model nor the three-point-source model can reproduce the H band reconstructions.
Both models have difficulty matching the $\mathrm{K_s}$ morphology as well, with the three-point-source model providing a slightly better match. 
In these cases, the amount of Gaussian noise that must be added to the simulation to match the scatter in the real data (Section \ref{sec:imrecon_mod}) is relatively large compared to the model observables themselves.
This is consistent with the Strehl, and thus data quality, decreasing from L$^\prime$ to H band.
The equally poor reconstructions at H band are also in agreement with the nearly identical $\chi^2$ values for the polar Gaussian ring and three-point-source models (Table \ref{tab:modelsel}).

In Appendix \ref{sec:app_recons} we further explore the effects of signal-to-noise and Fourier coverage on the image reconstructions.
Those simulations show that while the $\mathrm{K_s}$ and $\mathrm{H}$ observations have significantly lower signal-to-noise than the L$^\prime$ data, the reconstructed images are not consistent with the null model.
Rather, they display a significant asymmetry to the northeast of the star that (based on the $\chi^2$ values in Table \ref{tab:modelsel}) can be described equally well by either a Gaussian ring or multiple-point-source morphology at H, and slightly better by a multiple-point-source morphology at K$\mathrm{_s}$.
Also in Appendix A, we perform additional simulations to illustrate the locations in the images that are affected by the outer Gaussian ring (which resides approximately at the location of LkCa 15's outer, $\sim50$ AU disk).

\section{Discussion}\label{sec:disc}
The geometric modeling results presented in Sections \ref{sec:geomres}-\ref{sec:imreconsims} allow us to explore LkCa 15's morphology in a controlled way.
The two model classes (multiple point source and polar Gaussian ring) enable comparisons of point-like, ``clumpy" models to smooth, extended models. 
They also enable searches for variability via goodness-of-fit statistics, geometric parameter estimations, and image reconstruction tests. 
These tests have thus far demonstrated that the data prefer the dynamic, three-point-source geometric model over the others. 
This suggests that the observations support a ``clumpy" source morphology over a smooth one, and that the multi-epoch datasets show evidence for significant variability from epoch to epoch. 

Since these models are simple, analytic parameterizations of the source morphology, the fit results could be interpreted in multiple ways from a physical standpoint. 
For example, at the angular scales probed here, a dynamic multiple-point-source model could be used to describe: (1) self-luminous orbiting companions such as accreting protoplanets, (2) forward scattered light by massive companions shrouded in dust , (3) forward scattering by complex azimuthal disk asymmetries at $\sim20$ AU, or (4) rapidly-varying shadowing of the $\sim20$ AU forward-scattering disk by close-in circumstellar material. 
In the following sections we discuss these possible physical interpretations given the modeling and image reconstruction results.
We then discuss the characteristics of the solar-system-scale regions of the LkCa 15 system based on this interpretation.

\subsection{Three Self-Luminous Companions Cannot Reproduce the Data}

One possible explanation for a time-variable, three-point-source geometry is the presence of three orbiting, self-luminous companions (such as accreting protoplanets). 
This interpretation was applied to the 2009-2016 Keck and LBT datasets.
\citet{2015Natur.527..342S} fit Keplerian orbits in the plane of the outer disk to the 2009-2010 three-point-source astrometry published in \citet{2012ApJ...745....5K} and the 2014-2015 $\mathrm{K_s}$ and L$^\prime$ LBT astrometry.
They found that the three point sources (LkCa 15 b, c, and d) appeared to be on distinct orbits with semimajor axes between 14 and 20 AU. 
\citet{2016SPIE.9907E..0DS} added the two 2016 L$^\prime$ epochs (with just two companions identified as LkCa 15 c and d), and showed them to be consistent with the published orbital constraints. 

\subsubsection{Comparison to Orbital Motion Expectations}
The simplest physical test of whether this explanation holds given the longer time baseline is whether the three-point-source fits ever result in position angles away from the forward-scattering side of the protoplanetary disk. 
Orbiting, self-luminous companions should have position angles ranging from 0$^\circ$ to 360$^\circ$, while spurious companion signals caused by forward scattering will have position angles confined to one side of the disk.
Here we examine the multi-epoch three-point-source position angles in this context, and also compare them to expectations for orbital motion.

\begin{figure*}
\begin{centering}
\includegraphics[width=0.95\textwidth]{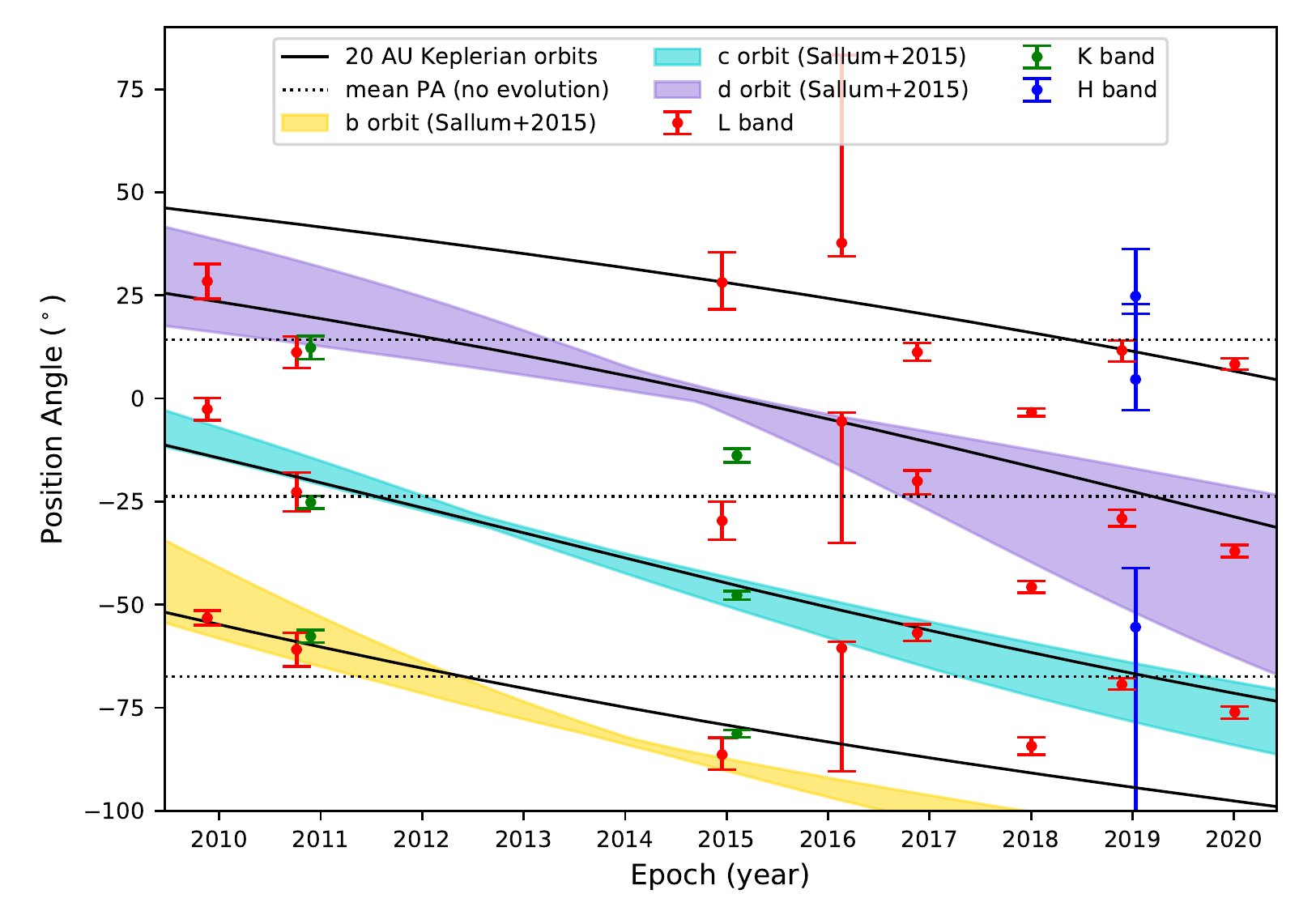}
\caption{Position angle evolution of three-point-source model fits compared to orbiting and stationary scenarios. Scattered points with error bars show best-fit position angles of three-point-source models fit to each L band (red), K band (green), and H band (blue) epoch. The first three sets of points show the fit results published in \citet{2012ApJ...745....5K}, which we include since they were used for the orbital fitting published in \citet{2015Natur.527..342S}. Gold, turqoise, and purple shaded curves show the best-fit orbits to the LkCa 15 b, c, and d candidates identified in \citet{2015Natur.527..342S}, respectively. Horizontal dotted lines indicate constant position angles. Solid curves show $\sim$20 AU Keplerian orbits aligned with the outer disk (Section \ref{sec:orbfits}). As discussed in Section \ref{sec:orbfits}, although we only fit three-point-source geometric models at any given epoch, we include the four Keplerian orbits to illustrate the expected signal for forward scattering by clumps orbiting through the disk. \label{fig:comp_PA_ev}}
\end{centering}
\end{figure*}

\begin{figure*}
\begin{centering}
\includegraphics[width=0.9\textwidth]{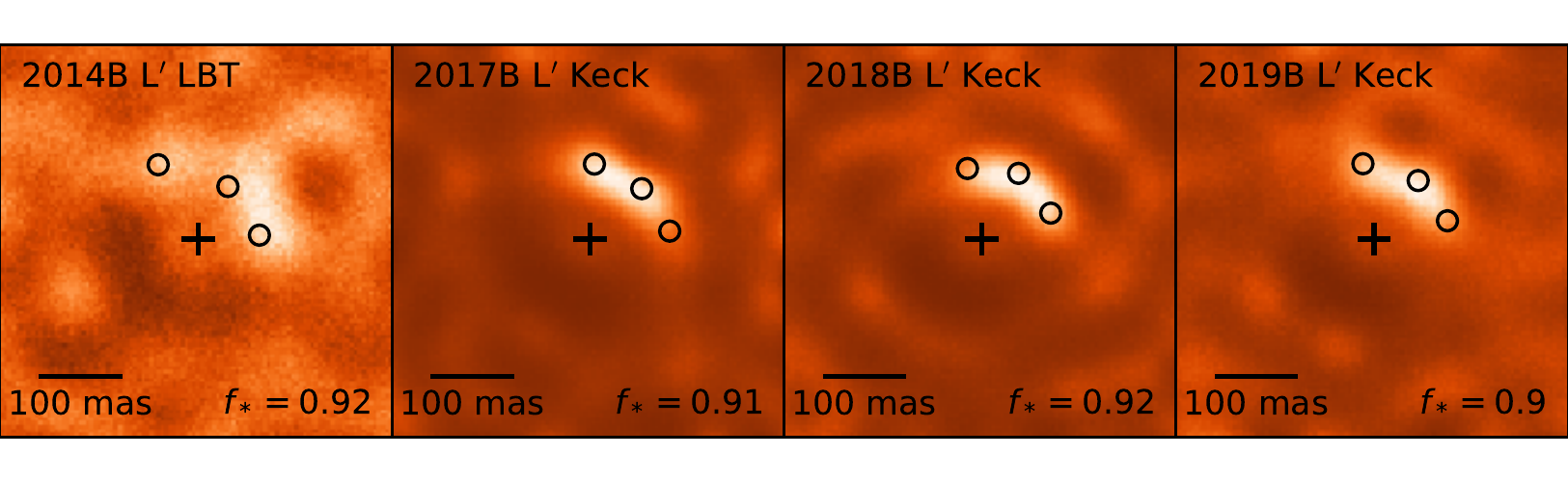}
\caption{Images show example L$^\prime$ reconstructed images with hollow circles indicating the best-fit three-point-source model positions. The position angles of the circles in each panel correspond to the red points in Figure \ref{fig:comp_PA_ev}. The three-point-source model positions generally overlap with bright regions in all of the reconstructions. However, since the model requires all off-axis flux to be concentrated in three delta functions, the point sources do not necessarily always fall on the brightest pixels. \label{fig:Lp_ims_comps}}
\end{centering}
\end{figure*}

Figure \ref{fig:comp_PA_ev} shows the best-fit three-point-source position angles for the 2009-2016 epochs, along with the four new Keck epochs from 2017-2020.
Figure \ref{fig:Lp_ims_comps} shows an example of L$^\prime$ three-point-source model fits plotted over the reconstructed images.
The 1$\sigma$ orbits published in \citet{2015Natur.527..342S} generally agree with the astrometry from the newly-reduced LBT data.
However, one notable difference for the 2016 epochs is that, when a three-point-source model is enforced, the predicted orbits for c and d overlap with the southwest and central sources, respectively (as opposed to northeast and central, which were identified as c and d in \citet{2015Natur.527..342S}).
The 2017-2020 Keck datasets yield the same results; the central and southwest sources overlap with the c and d orbital estimates, while a northeast source exists away from any orbital predictions. 
In short, once the predicted position of LkCa 15 b exits the forward scattering side of the disk, the best-fit point source position angles are inconsistent with orbital expectations.

This result is consistent with the morphology in the reconstructed images (Figure \ref{fig:recons_epochs}). 
The infrared emission stays on one side of the star, at the expected orientation for forward scattered light based on prior disk constraints \citep[e.g.][]{2016ApJ...828L..17T,2022ApJ...937L...1L}.
The three-point-source fit then yields position angles that are constant in time to within $\sim30-40^\circ$. 
This is in agreement with \citet{2019ApJ...877L...3C}, which presented follow-up observations with Keck/NIRC2 imaging and did not detect b at its predicted position angle based on the 2009-2015 orbit fitting.
The preference for the dynamic three-point-source model thus cannot imply the presence of three self-luminous point masses on distinct orbits, but may imply forward-scattered light.

\subsubsection{Comparison with the LkCa 15 b H$\alpha$ Detection}\label{sec:halpha}
While the multi-epoch L$^\prime$ datasets suggest that forward scattered light is the source of the NRM signals, the simultaneous detection of LkCa 15 b at H$\alpha$ in 2014 \citep[e.g.][]{2015Natur.527..342S,2022arXiv221102109F} is seemingly at odds with this hypothesis.
LkCa 15 b's H$\alpha$ luminosity implies emission in excess of expectations for forward scattered light, given its contrast in the nearby continuum and the H$\alpha$-to-continuum ratio of the star (which should be the same as the H$\alpha$-to-continuum ratio of forward scattered light).
The natural explanation for this excess is accretion, since infalling hydrogen gas should become shocked and exhibit line emission \citep[e.g.][]{2020arXiv201106608A,2021ApJ...917L..30A,2022RNAAS...6..262M}.
Here we present a new method for estimating the signal-to-noise ratio of accretion signals detected in the vicinity of forward-scattered light.
We use this method along with H$\alpha$ measurements of LkCa 15 b taken from the literature to explore whether the H$\alpha$ detection could be consistent with the inner disk scenario.

The previously-published H$\alpha$ detection was made via a spectral-differential imaging (SDI) reduction of Magellan/MagAO data obtained simultaneously at H$\alpha$ and in the adjacent continuum \citep[e.g][]{2014ApJ...781L..30C}.
Two different data reduction approaches are taken depending on the likelihood of forward scattering at the location of a signal of interest.
In the first, continuum images are simply subtracted from H$\alpha$ images to perform the SDI step.
In the second, continuum images are first scaled by the ratio of the stellar fluxes in H$\alpha$ and continuum (measured via aperture photometry), before being subtracted from H$\alpha$ observations \citep[e.g.][]{2022arXiv221102109F}.

The reasoning behind the latter approach is that forward scattered light will display an excess at H$\alpha$ (relative to the continuum) if the central star has an H$\alpha$ excess.
By scaling the continuum dataset before subtraction, any forward scattering signals should be eliminated, since the H$\alpha$-to-continuum ratio in scattered light should be the same as the stellar ratio. 
Describing this in mathematical terms, in the presence of forward scattering by dust, the H$\alpha$ excess measured via SDI ($H\alpha_{ex}$) is equal to:
\begin{equation}\label{eq:scaled_sdi}
    H\alpha_{ex} = H\alpha_{tot} - C_{tot} \frac{H\alpha_{*}}{C_{*}},
\end{equation}
where $H\alpha_{tot}$ is the total $H\alpha$ signal at a given location, assumed to be a combination of accretion and forward scattered light ($H\alpha_{tot} = H\alpha_{a} + H\alpha_{s}$). 
The parameter $C_{tot}$ is the total continuum signal at the same location, assumed to be purely forward scattered light, and $H\alpha_{*}$ and $C_{*}$ are the H$\alpha$ and continuum fluxes, respectively, measured for the star.
In this conservative SDI approach, any H$\alpha$ signals with contributions due to forward scattering will have lower significance when Equation \ref{eq:scaled_sdi} is applied.

LkCa 15's stellar H$\alpha$-to-continuum ratio was measured at 1.81$\pm$0.03 in the 2014 observations \citep{2022arXiv221102109F}. 
When the conservative SDI approach is applied, the H$\alpha$ signal-to-noise estimate decreases from 4.9 (in the version with no scaling prior to subtraction) to 2.9 \citep{2022arXiv221102109F}, with the unscaled version closer to the original signal-to-noise published in \citet{2015Natur.527..342S}. 
The lower significance of the conservative-SDI reduction suggests that scattering from the inner disk at least contributed to the H$\alpha$ excess.

Beyond these lower significance values, it is also worth carefully considering the additional noise contributions at the location of the excess when forward scattering is present.
In the framework of Equation \ref{eq:scaled_sdi}, the $H\alpha_{ex}$ term is on-average equal to $H\alpha_{a}$ if $C_{tot}$ is equal to the forward scattered light in the continuum ($C_{s}$). 
The propagated noise at the location of the excess ($\sigma^2_{H\alpha,ex}$) is then:
\begin{equation}\label{eq:scat_sdi}
    \sigma^2_{H\alpha_{ex}} = \sigma^2_{H\alpha_a} + \sigma^2_{H\alpha_s} + \sigma^2_{C_s} f^2_{*} + C^2_{s} \sigma^2_{f_*},
\end{equation}
where $\sigma_{H\alpha_{a}}$ and $\sigma_{H\alpha_s}$ are noise contributions from the accretion signal and scattering signal at H$\alpha$, $\sigma_{C_s}$ is the noise contribution from the continuum forward-scattered light, $f_{*}$ is the stellar H$\alpha$-to-continuum ratio ($\frac{H\alpha_{*}}{C_{*}}$), and $\sigma_{f_*}$ is its uncertainty.
We can then define a signal-to-noise penalty as a multiplicative scaling from the no-scattering case to the scattering case, which is simply the ratio of the noise without scattering to the noise with scattering:
\begin{equation}\label{eq:snrpen}
\begin{split}
    p_{\mathrm{SNR}} & = \frac{\mathrm{SNR_{no~scattering}}}{\mathrm{SNR_{scattering}}} = \frac{\sigma_{H\alpha_{a}}}{\sigma_{H\alpha_{ex}}}\\ & = \frac{\sigma_{H\alpha_{a}}}{\sqrt{ \sigma^2_{H\alpha_a} + \sigma^2_{H\alpha_s} + \sigma^2_{C_s} f^2_{*} + C^2_{s} \sigma^2_{f_*}}} 
\end{split}
\end{equation}

It is useful to discuss the difference between applying these equations in the conservative and simple SDI regimes (with and without scaling continuum images by $f_*$, respectively). 
A simple SDI approach is well justified if the star is known to have no H$\alpha$ excess. 
In this case, $f_* = \frac{H\alpha_{*}}{C_{*}} = 1$, and $\sigma_{f_*} = 0$, eliminating the final terms in Equation \ref{eq:scat_sdi} and the Equation \ref{eq:snrpen} denominator.
The uncertainty in the SDI H$\alpha$ excess would then have contributions from accretion, scattered light at H$\alpha$, and scattered light in the continuum, with the latter two being equal.
If the conservative SDI were applied in this situation, it would only add noise to the SDI measurements (by including an estimate of $f_*=1$ with Poisson noise from the star in both bandpasses). 
However, since an unaccounted stellar H$\alpha$ excess would create false-positive signals in the presence of forward scattering, the simple SDI should only be applied if the star definitively has no H$\alpha$ excess.

\begin{figure}
    \centering
    \includegraphics{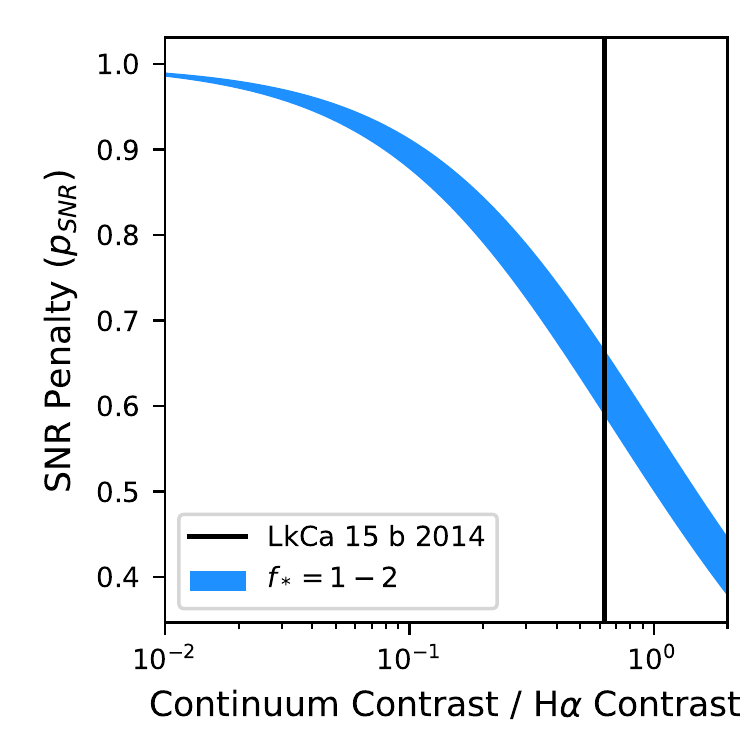}
    \caption{The shaded region shows the multiplicative signal-to-noise penalty (Equation \ref{eq:snrpen}) for an SDI-detected H$\alpha$ excess as a function of the contrast ratio of accretion-tracing H$\alpha$ and continuum forward scattered light, as well as the stellar H$\alpha$-to-continuum ratio ($f_*$; thickness of shaded region). The stellar H$\alpha$-to-continuum ratio was varied from 1 (no stellar accretion signature) to 2. The vertical line shows the continuum contrast to H$\alpha$ contrast ratio estimate for the 2014 LkCa 15 b detection (Section \ref{sec:halpha}). Because of forward-scattering contributions at H$\alpha$ and in the continuum image subtracted during the SDI process, performing SDI in the presence of forward-scattered light can suppress signal-to-noise significantly. This effect is not necessarily taken into account during conventional SDI signal-to-noise estimates.}
    \label{fig:noise}
\end{figure}

Since LkCa 15 and many young stars exhibit stellar H$\alpha$ excesses, we use Equation \ref{eq:snrpen} to illustrate how the SDI H$\alpha$ signal-to-noise value would be degraded by these extra noise terms assuming Poisson statistics and a conservative SDI approach.
For a continuum forward scattering signal with a contrast of $\sim5$ magnitudes ($C_s$ = 0.01 $C_*$), and a value for $f_*$ comparable to that for LkCa 15 in 2014 ($1.81\pm0.03$), the equal-contrast scattering contribution at H$\alpha$ is then $H\alpha_s = 0.01~H\alpha_* = 0.0181~C_*$.
Assuming we are looking for a similar contrast ($\sim5$ mag) H$\alpha$ accretion signature, we can calculate $H\alpha_a = 0.01~H\alpha_* = 0.0181~C_*$.
For Poisson statistics the noise terms can be written as the square root of the fluxes (e.g. $\sigma_{H\alpha_{a}} = \sqrt{0.0181~C_*}$). 
Plugging in the above numbers appropriately, the penalty is 0.52, meaning that the signal-to-noise value is nearly halved in the presence of an equal-brightness forward scattering signal.
Figure \ref{fig:noise} illustrates this effect for a range of forward scattering contributions relative to accretion contributions, as well as $f_*$ values. 

The non-SDI H$\alpha$ and continuum images from \citet{2022arXiv221102109F} show emission at the position of LkCa 15 b at both bands.
If we use the reduction in signal-to-noise between the non-scaled SDI and scaled SDI (from 4.9 to 2.9) to estimate their relative fluxes, assuming Poisson noise statistics results in an H$\alpha$-to-continuum flux ratio of 2.85. 
Accounting for the stellar H$\alpha$-to-continuum ratio, the continuum contrast is then the H$\alpha$ contrast divided by 2.85 (lower companion flux in continuum) times 1.81 (lower stellar flux in continuum), or 3.9 mag for the H$\alpha$ contrast of 3.4 mag measured in \citet{2022arXiv221102109F}.
This implies that the signal-to-noise estimate for the 2014 LkCa 15 b detection should be scaled down by a factor of $\sim0.67$ if these noise terms are not taken into account (Figure \ref{fig:noise} vertical line).

Signal-to-noise estimators that compare the SDI excess flux to fluxes in an annuli at the same radius do not adequately take scattering noise sources into account. 
Forward scattered light will not be present in the entire comparison annulus, which would bias the empirically measured noise toward low values. 
This type of estimator was applied in \citet{2015Natur.527..342S}.
Bayesian approaches applied more recently  \citep{2022arXiv221102109F}, such as the \texttt{pyklip} \texttt{PlanetEvidence} module, can also be biased by these local noise spikes \citep[e.g.][]{2019arXiv191201232G}.
Combining the lower signal-to-noise values from the re-reduction presented in \citet{2022arXiv221102109F} with these additional signal-to-noise concerns, a conservative estimate of the LkCa 15 b H$\alpha$ signal to noise is then 1.9. 
The LkCa 15 b H$\alpha$ detection is thus not strongly inconsistent with the forward scattering scenario.

\subsection{The LkCa 15 Disk on Solar System Scales}
\subsubsection{A Dynamic Disk Morphology}\label{sec:orbfits}
Given the lack of coherent orbital motion, the proximity of the geometric model components to known locations of forward-scattered light \citep[e.g.][]{2016ApJ...828L..17T}, and the decreased H$\alpha$ detection significance, forward scattering by disk material provides the best explanation for the multi-epoch data.
While orbital motion is not observed beyond the forward-scattering side of the disk, the preference for the dynamic geometric models suggests that the forward-scattering signal must vary with time. 
Such a dynamic signal could be caused by azimuthal asymmetries and/or dust-shrouded companions orbiting through the disk, or by shadowing due to material closer to the star.

The dust-shrouded companions hypothesis is unlikely, given the need for multiple massive companions at the same orbital semimajor axis, which would be dynamically unstable \citep[e.g.][]{1993Icar..106..247G}. 
This leaves a dynamic disk as the best explanation for the observations. 
The variability could be explained by aliasing of quickly-changing shadowing by closer-in disk material.
This type of shadowing has been seen in systems similar to LkCa 15 \citep[e.g.][]{2016A&A...595A.113S}.
Furthermore, polarized light observations of LkCa 15 presented in \citet{2016ApJ...828L..17T} revealed azimuthal asymmetries in the outer, $\sim50$ AU ring that suggested shadowing, although it was not clear whether this was caused by the $\sim20$ AU disk or interior material. 
The evidence for a sub-AU disk from spectral energy distribution fitting \citep{2007ApJ...670L.135E}, and the recent detection of a $\sim5$ AU disk \citep{2022ApJ...931....3B} suggest that the latter scenario is possible, and lend support to the shadowing explanation for these NRM observations.

As an alternative explanation, complex and variable disk features at the locations of the infrared sources would be consistent with ALMA observations of the outer regions of the disk, which resolve azimuthal overdensities in the form of clumps and arcs \citep{2022ApJ...937L...1L}.
To explore this scenario, we return to the position angles in Figure \ref{fig:comp_PA_ev}, which show significant variation confined to the forward-scattering side of the star. 
This behavior would be expected as azimuthal overdensities orbit through the disk.
In this scenario, companion-like signals would appear on one edge of the forward scattering arc and exhibit position angle evolution until they orbit out of the forward scattering region. 
The position angle evolution in Figure \ref{fig:comp_PA_ev} is generally consistent with this scenario. 
For example, the signal identified as LkCa 15 b has a position angle that decreases from $\sim-50^\circ$ to $\sim-80^\circ$ from 2014 to 2016. 
In late 2017 no signal is present at the expected position of b, and a new signal exists to the northeast of the predicted c and d positions. 

We test whether a simple multiple-clump case can reproduce the data by fitting four orbits to the position angles in Figure \ref{fig:comp_PA_ev}.
We explore Keplerian orbits aligned with the millimeter disk, fixing the semimajor axes to $20$ AU and allowing the true anomalies to vary. 
Figure \ref{fig:comp_PA_ev} (solid lines) shows the results, which roughly match the position angle evolution of the  geometric model. 
The orbital fit provides a better match to the data than static position angle models, with a $\chi^2$ improvement of 55 relative to three static position angles, and an improvement of 4 relative to four static position angles. 
However, all of the $\chi^2$ values are high ($\sim1075-1130$ for $28-29$ degrees of freedom), showing that none of these models are a good fit to the data. 
These tests, while simple, suggest that a plausible explanation for the multi-epoch, multi-companion fits is that they resolve complex azimuthal brightness variations in a dynamic disk, as opposed to something like a small number of orbiting clumps.

Higher cadence observations could help to more firmly constrain this type of variability and distinguish between these physical explanations, especially since shadowing by inner disk material has been shown to evolve on timescales of $\sim$days \citep[e.g.][]{2018ApJ...868...85P}.  
While several of the observational epochs presented here consist of two adjacent nights, they are combined in order to maximize signal to noise while probing month-to-year-long variability.
For all multi-night epochs, poor conditions and/or limited sky rotation during at least one of the nights precludes a search for variability on nightly timescales.
To explore whether high-quality, closely-spaced nights could accomplish this, we compare the constraints from the 2018-01 L$^\prime$ epoch (which consists of the 171231 and 180101 nights) and the 180101 night alone.
This test shows that the 180101 night is of sufficient quality to constrain the geometric parameters.
Thus, with sufficient data quality, future Keck (or similar) NRM studies carried out at multiple cadences (e.g. from nightly to annual) could better distinguish between orbiting azimuthal asymmetries and fast-varying shadowing.

\subsubsection{Multi-Wavelength Disk Properties}
The H band data presented here are the first detection of the LkCa 15 inner disk at that wavelength.
Previous H band NRM observations at Keck resulted in a non-detection and estimated the lower limit on the contrast of the disk to be 7 magnitudes per resolution element \citep{2014IAUS..299..199I}. 
The multiple point source geometric fit results suggest that the 2019-01 H band Keck data are more sensitive, since the point source contrasts (which range from 6.95-7.24 mag) effectively correspond to detections in individual resolution elements.
The polar Gaussian ring fit results (Table \ref{tab:diskfits}) suggest that the integrated contrast of the inner ring (and thus the integrated contrast of forward scattering at that location) is 3.8 magnitudes. 
This is consistent with the three-point-source fit, since the extent of the inner ring model component is roughly 15 resolution elements at H band. 

Together with the $\mathrm{K_s}$ and $\mathrm{L^\prime}$ observations we can place multi-wavelength constraints on the disk properties. 
We can use the framework of the polar Gaussian ring models to understand the geometric properties of the small-grain disk ring at $\sim$20 AU.
For those models, the spatial properties (e.g. radius, width, and inclination) of the inner ring are constant with wavelength to within 1-2$\sigma$.
Notably, the fractional fluxes are also constant within the 1$\sigma$ errors, suggesting that the properties of the small grain dust lead to forward scattering with relatively flat spectral slope.
This is generally consistent with the disk properties estimated from the K and L$^\prime$ fluxes in \citet{2019ApJ...877L...3C}, which posited that a large ($\gtrsim0.5~\mu$m) minimum grain size is required to match the disk's infrared colors.
Future studies utilizing radiative transfer modeling could use the H band data presented here to further constrain the dust grain distribution.  

The outer ring parameter constraints from the polar Gaussian ring models can be used to understand the geometry of LkCa 15's outer disk (which has a spatial scale of $\sim50$ AU).  
This model component is most radially-compact at $\mathrm{K_s}$, most azimuthally-compact at H, and most symmetric at L$^\prime$ (Figure \ref{fig:2disk_bands} and Table \ref{tab:diskfits}). 
Of the three, the L$^\prime$ geometry is most similar to the constraints from \citet{2022ApJ...931....3B}. 
The fractional fluxes for the outer ring also differ across the bands, with the highest flux at L$^\prime$, followed by $\mathrm{K_s}$ and then H.
Within the assumptions of the model, the outer ring morphological differences from band to band appear to be more significant ($2-5\sigma$) than those for the inner ring.
However, as described below, the outer ring modeling has a few caveats that prevent us from making strong conclusions based on these differences.

One aspect of the $\sim50$ AU disk in LkCa 15 that is not explicitly captured by the model is an apparent offset from the star of several tens of milliarcseconds \citep[e.g.][]{2014A&A...566A..51T}, which may be caused by dynamical shaping by an unseen companion.
Since the combined L$^\prime$ dataset has the highest signal to noise and the best sensitivity to the outer ring (see Appendix \ref{sec:app_recons}), we test whether allowing for ring offsets significantly changes the brightness distribution of the best-fit model.
The results show that the best-fit offset ring parameters change to put both the inner and outer ring peak flux in locations identical to the centered ring model. 
The only notable difference is that the offset outer ring's faint side (to the southeast) is located slightly closer to the central star than it is for the centered outer ring.

The offset inner ring is virtually identical to the centered inner ring, which is unsurprising given that its best-fit offset is consistent with zero at just above 1$\sigma$. 
We thus conclude that the bright side of the rings in the best-fit centered models are consistent with an offset outer disk in LkCa 15, but that a physically-motivated model (produced by e.g. hydrodynamic and radiative transfer simulations) may lead to a slightly different morphology, especially for the fainter side of the disk. 
Given that the centered and offset inner rings are nearly identical, this limitation in the modeling is not an issue for its characterization, and for identifying epoch-to-epoch variability.

In addition to the above caveats for the outer disk constraints, the incomplete Fourier sampling and the large outer ring angular scales make its morphological constraints poorer than the inner ring.
This is especially true for the shortest H and $\mathrm{K_s}$ wavelengths that may alias spatial frequencies that probe the outer disk, and that have poorer signal-to-noise than the L$^\prime$ datasets (as evidenced by the larger reduced $\chi^2$ values in Table \ref{tab:modelsel}, and the image reconstruction simulations in Appendix \ref{sec:app_recons}).
These limitations may make conventional imaging methods \citep[such as coronagraphy and/or reference differential imaging;][]{2019AJ....157..118R} more useful for constraining its properties.

\section{Conclusions}\label{sec:conc}
We presented the highest angular resolution infrared monitoring of the LkCa 15 system using Keck and LBT non-redundant masking inteferometry.
We generated reconstructed images from, and performed geometric model fits to these multi-epoch and multi-wavelength data, with the following findings:
\begin{enumerate}[leftmargin=*]
    \item Fitting the multi-epoch observations with simple geometric models shows that the data strongly prefer a clumpy source morphology to a smooth one, and a dynamic scenario to a static one. This is consistent with the image reconstructions, which show structure that is best matched by a multiple-point-source geometric model and variability that can only be reproduced by dynamic geometric models.
    \item The preference for the multiple-point-source geometric model cannot be physically explained by three self-luminous, orbiting companions (such as accreting protoplanets). The position angles of the best-fit model components and the asymmetries in the reconstructed images are confined to the forward-scattering side of the protoplanetary disk, making scattered light by disk material a natural explanation for the infrared signals.
    \item We quantify additional noise contributions to H$\alpha$ SDI performed in the presence of forward scattered light, which are not taken into account during conventional signal-to-noise estimates for SDI H$\alpha$ excesses. These noise considerations show that the H$\alpha$ detection cannot strongly rule out the forward scattering scenario, making it a plausible explanation for the H$\alpha$ detection in addition to the infrared sources.
    \item The variations in the geometric modeling and image reconstructions indicate changing azimuthal asymmetries on $\sim20$ AU scales. These may be caused by overdensities orbiting through the forward scattering side of the disk or fast-varying shadowing by closer-in disk material. Both of these scenarios would be more easily described by a multiple point source model than a smooth polar Gaussian ring model, in agreement with the preference for the former in fits to the data.
    \item We make the first detection of the inner disk at H band, with a geometry that is consistent with previous K and L band studies (with both NRM and conventional imaging). 
    \item The geometric fits to the multi-wavelength data suggest that the dust properties at $\sim$20 AU must lead to a relatively flat spectral slope. This is in agreement with previous K and L band imaging. More sophisticated radiative transfer simulations could better connect the H band observables to quantitative disk and dust grain properties.
\end{enumerate}

These data demonstrate the utility of long time baseline monitoring of young stellar systems at high angular resolution. 
The early interpretations of LkCa 15 NRM data in \citet{2012ApJ...745....5K} and \citet{2015Natur.527..342S} were based on short time baseline observations where disk signals would more easily masquerade as orbiting point masses.
The longer lever arm created by the additional Keck epochs clearly shows that any point-like signals cannot be caused by three individual point masses.
\citet{2012ApJ...745....5K} and \citet{2015Natur.527..342S} also lacked careful simulation techniques for understanding NRM systematics in e.g.~image reconstruction, which have been more thoroughly developed in the intervening years. 
Despite these shortcomings, the early Keck NRM data led to the discovery of LkCa 15's dynamic small-grain disk several years earlier than it was characterized with conventional imaging techniques \citep[e.g.][]{2016ApJ...828L..17T,2019ApJ...877L...3C}.
This highlights the value of NRM for making early observations on extreme angular scales, especially when equipped with modeling approaches that enable robust interpretation of complex signals.

The high angular resolution offered by NRM, and the time resolution of the multi-epoch datasets allow us to place the first constraints on dynamic small-grain disk substructures in LkCa 15.
Future higher cadence monitoring of this system could enable high-fidelity image reconstructions at more closely-spaced epochs. 
This would place better constraints on the details of the disk variability, perhaps distinguishing between shadowing and changing azimuthal asymmetries via time-resolved mapping.
Interferometric techniques applied on the next generation of 30-meter telescopes will enable similar studies with even greater angular resolution, revealing the dynamics of disk material down to $\sim$few AU orbits where the vast majority of mature giant planets have been found.

\begin{acknowledgements}
SS acknowledges support from the National Science Foundation under Grant No. 2009698. 
JAE acknowledges support from the National Science Foundation under Grant No. 1745406.
Some of the data presented herein were obtained at the W. M. Keck Observatory, which is operated as a scientific partnership among the California Institute of Technology, the University of California and the National Aeronautics and Space Administration. The Observatory was made possible by the generous financial support of the W. M. Keck Foundation. The authors wish to recognize and acknowledge the very significant cultural role and reverence that the summit of Maunakea has always had within the indigenous Hawaiian community.  We are most fortunate to have the opportunity to conduct observations from this mountain.
\end{acknowledgements}

\begin{appendix}
\section{Additional Image Reconstruction Tests}\label{sec:app_recons}
While the constraints presented in this work are based on model fits to the Fourier observables, high-quality image reconstructions may provide more detailed information for complex sources that are not well fit by simple geometric models. 
Here we perform additional image reconstruction simulations to explore the fidelity of the SQUEEZE images.

\begin{figure}
\begin{center}
\includegraphics[width=\columnwidth]{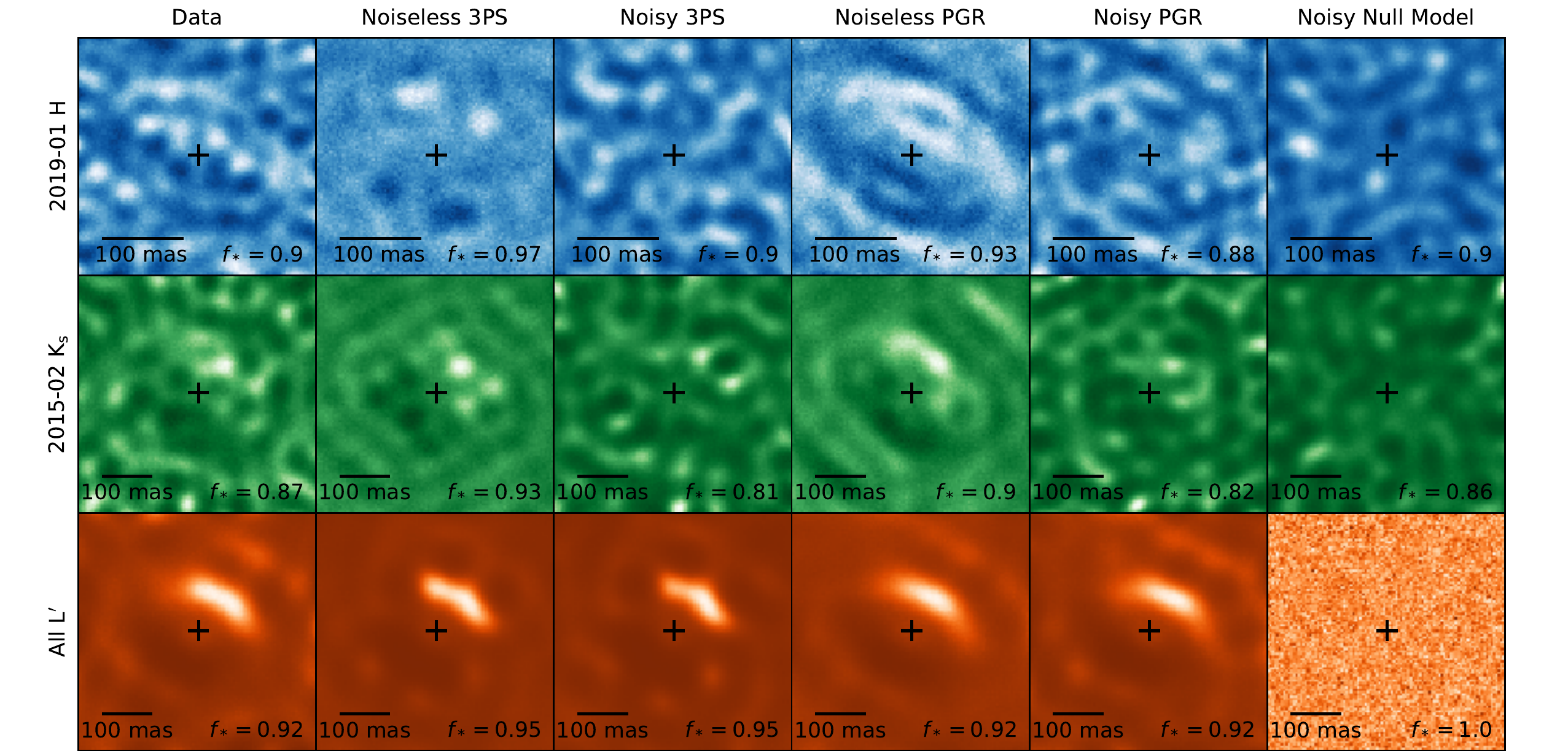}
\end{center}
\caption{Comparison of observed reconstructed images in combined datasets from each band to simulated reconstructions of best-fit models for the same datasets. Columns from left to right show images reconstructed from: (1) combined datasets from each band, (2) simulated noiseless observations of the best-fit three-point-source model, (3) simulated noisy observations of the best-fit three-point-source model, (4) simulated noiseless observations of the best-fit Gaussian ring model, (5) simulated noisy observations of the best-fit Gaussian ring model, and (6) simulated observations of the null, point-source model, with noise added to match the scatter in the real data. From top to bottom, rows correspond to the 2019-01 H band, 2015-02 $\mathrm{K_s}$ band, and 2014-2020 combined L$^\prime$ datasets. Comparison of the noiseless and noisy reconstructions illustrates the different effects that signal to noise ratio and Fourier coverage have on imaging quality. Comparison of the images reconstructed from the data to those in column (6) illustrates (qualitatively) the significance of the detections given the systematic effects of SQUEEZE and the signal-to-noise of each dataset. The reconstructions in columns (1), (3), and (5) are identical to those shown in Figure \ref{fig:comb_disk_comp_recons}. 
\label{fig:comb_disk_comp_recons_noisetest}}
\end{figure}

We first carry out tests to assess the relative contributions of signal-to-noise ratio and Fourier coverage to systematics in the images, by simulating observations of different geometric models with and without added noise (indicated by the ``Noiseless" and ``Noisy" labels in Figure \ref{fig:comb_disk_comp_recons_noisetest}). 
For the noiseless reconstructions, we simulate mock observations with Fourier coverage identical to each dataset, but with no noise added to the closure phases and squared visibilities.
The noisy reconstructions include Gaussian noise added to the Fourier observables, at a level that causes the standard deviation of the simulated closure phases and squared visibilities to equal that of the real data (as described in Section \ref{sec:imrecon_mod}). 
For all reconstructions, we preserve the relative weighting of the various baselines and closing triangles by keeping the assigned error bars equal to those of the real data.

Figure \ref{fig:comb_disk_comp_recons_noisetest} shows the results for simulated observations of the best-fit three-point-source and Gaussian ring models for the combined H, $\mathrm{K_s}$, and L$^\prime$ datasets. 
Comparison of the noiseless Gaussian ring images (Figure \ref{fig:comb_disk_comp_recons_noisetest} column 4) to the geometric models shown in Figure \ref{fig:2disk_bands} shows that the Fourier coverage and baseline weighting alone can bias the reconstructions. 
This is especially the case when imaging extended structures, which has been explored more thoroughly in \citet{2017ApJS..233....9S}.
Comparing the noiseless and noisy reconstructions shows that degrading the signal-to-noise ratio can increase the degree to which extended features are over-resolved (columns 4 versus 5), and can also decrease the significance of point source detections (columns 2 versus 3).

Aside from generally demonstrating the effects of noise and Fourier coverage on the SQUEEZE images, Figure \ref{fig:comb_disk_comp_recons_noisetest} shows that the L$^\prime$ dataset has the highest signal to noise. 
There is very little difference between the noiseless and noisy L$^\prime$ images, with the exception of the outer arc of emission to the northwest (up and right) of the star being slightly more overresolved into clumpy structures when noise is added. 
Lastly, comparing column 6 to column 1 shows that the images reconstructed from the observations are significantly different from the null model (an unresolved star).

\begin{figure}
\begin{center}
\includegraphics[width=0.7\columnwidth]{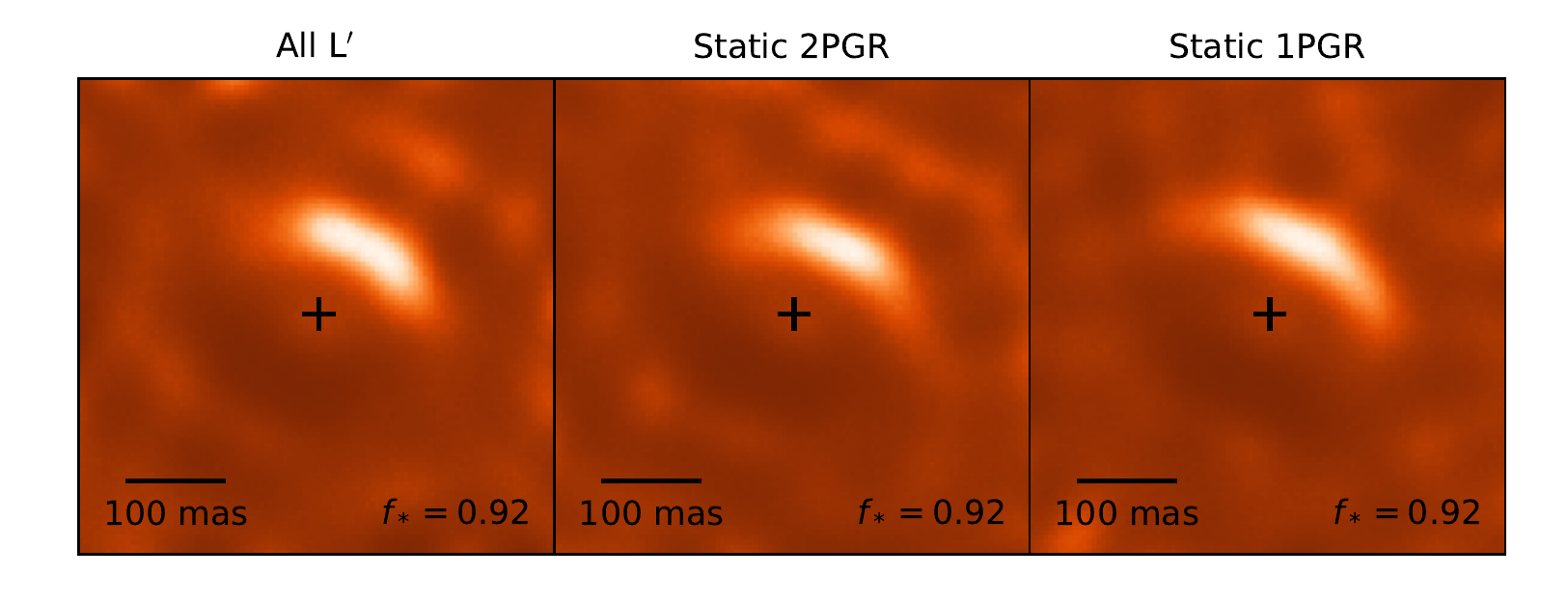}
\end{center}
\caption{Comparison of the observed reconstructed image for the combined L$^\prime$ dataset (left) to simulated reconstructions of best-fit one-ring (right) and two-ring (center) models. The two-ring scenario results in a pronounced outer arc to the northwest of the star, at approximately twice the separation of the bright inner arc, and at the same location where an outer arc is seen in the data. The two-ring model also results in slightly more pronounced emission to the southeast of the star at a similar angular separation. It is worth noting that the combined L$^\prime$ dataset has the highest sensitivity to this outer ring component (which roughly coincides with LkCa 15's $\sim50$ AU disk), so individual epochs may not show such noticeable changes in the image reconstructions between the two scenarios.
\label{fig:recons_ringtest}}
\end{figure}

We also simulate reconstructions of Gaussian ring models with and without an outer ring component to evaluate the sensitivity of the imaging to LkCa 15's $\sim50$ AU disk.
We follow the same process described in Section \ref{sec:imrecon_mod}, reconstructing a combined L$^\prime$ image for the best-fit models including one and two rings, respectively. 
Figure \ref{fig:recons_ringtest} shows the results.
Comparing the center and right panels shows that the outer ring causes an arc to the northwest of the star, at a location consistent with the outer arc in the All L$^\prime$ reconstruction. 
It also results in slightly more pronounced emission to the southeast of the star, which is consistent with the real data as well. 
Thus, the combined L$^\prime$ dataset is sensitive to the $\sim50$ AU scales where the outer ring model component exists, and where LkCa 15 is known to have an outer disk. 
However, the sparse Fourier coverage of individual epochs means that they may not all be so sensitive to these scales.
For example, the 2016-02 L$^\prime$ and 2016-11 L$^\prime$ reconstructions shown in Figure \ref{fig:recons_epochs} do not exhibit such a clear outer arc, but some of the later Keck epochs do, with 2018-11 L$^\prime$ having possibly the clearest signal at that location.

\end{appendix}

\bibliography{references2}{}
\bibliographystyle{aasjournal}

\end{document}